%% file: paper.tex
\documentclass[usenatbib,letterpaper,hyperref]{mn2e}
\usepackage[labelfont=bf,labelsep=period]{caption}
\usepackage[totalwidth=480pt, totalheight=680pt]{geometry}
\usepackage{hyperref}
\usepackage{times}
\usepackage{epsfig}
\usepackage{amsmath}
\usepackage{amssymb}
\usepackage{url}
\usepackage{aas_macros}
\def\eqsim{\mathrel{\raise0.35ex\hbox{$\scriptstyle =$}\kern-0.6em
    \lower0.40ex\hbox{{$\scriptstyle \sim$}}}}
\def\gtrsim{\mathrel{\raise0.35ex\hbox{$\scriptstyle >$}\kern-0.6em
    \lower0.40ex\hbox{{$\scriptstyle \sim$}}}}
\def\lesssim{\mathrel{\raise0.35ex\hbox{$\scriptstyle <$}\kern-0.6em
    \lower0.40ex\hbox{{$\scriptstyle \sim$}}}}
\def\HI{H\,{\sc i}}
\def\AA{ALFALFA}
\def\HIWF{H\,{\sc i}\,WF}
\def\HIMF{H\,{\sc i}\,MF}
\def\HIVF{H\,{\sc i}\,VF}

\defcitealias{2015AnA...574A.113P}{P15}
\defcitealias{2018MNRAS.477....2J}{J18}
\defcitealias{2011AJ....142..170H}{H11}

\title[The \AA\ {\rm H}\,{\sc i}\,WF]{The \AA\ \HI\ velocity width function}
\author[K. A. Oman]{
  Kyle A. Oman$^{1,2}$\thanks{kyle.a.oman@durham.ac.uk}\\
  $^{1}$ Institute for Computational Cosmology, Durham University, South Road, Durham DH1 3LE, United Kingdom\\
  $^{2}$ Department of Physics, Durham University, South Road, Durham DH1 3LE, United Kingdom\\
}
\date{\today}

\begin{document}
\label{firstpage}
\maketitle

\begin{abstract}
We make the most precise determination to date of the number density of extragalactic $21$-cm radio sources as a function of their spectral line widths -- the \HI\ velocity width function (\HIWF) -- based on $21827$ sources from the final $7000\,\mathrm{deg}^2$ data release of the Arecibo Legacy Fast ALFA ({\AA}) survey. The number density of sources as a function of their neutral hydrogen masses -- the \HI\ mass function (\HIMF) -- has previously been reported to have a significantly different low-mass slope and `knee mass' in the two sky regions surveyed during \AA. In contrast with this, we find that the shape of the \HIWF\ in the same two sky regions is remarkably similar, consistent with being identical within the confidence intervals implied by the data (but the overall normalisation differs). The spatial uniformity of the \HIWF\ implies that it is likely a stable tracer of the mass function of dark matter haloes, in spite of the environmental processes to which the measured variation in the \HIMF\ are attributed, at least for galaxies containing enough neutral hydrogen to be detected. This insensitivity of the \HIWF\ to galaxy formation and evolution can be exploited to turn it into a powerful constraint on cosmological models as future surveys yield increasingly precise measurements. We also report on the possible influence of a previously overlooked systematic error affecting the \HIWF, which may plausibly see its low-velocity slope steepen by $\sim 40$~per~cent in analyses of future, deeper surveys. Finally, we provide an updated estimate of the \AA\ completeness limit.
\end{abstract}

\section{Introduction}
\label{SecIntro}

One of the fundamental predictions of the current concordance cold dark matter ($\Lambda$CDM) cosmology is the number density as a function of mass of self-bound dark matter haloes, termed the halo mass function \citep[HMF;][]{1988ApJ...327..507F,2001MNRAS.321..372J,2002MNRAS.329...61S}. Once the parameters of the $\Lambda$CDM model are specified, the HMF in the absence of galaxy formation, e.g. as realized in a cosmological N-body simulation, is a prediction with no free parameters. The CDM HMF is well-approximated as a power law with a slope $\phi(M)\propto M^{-1.9}$ over $\gtrsim 8$ orders of magnitude below the cutoff at the scale of the largest collapsed structures at the present day \citep{2012MNRAS.426.2046A}. In a more realistic scenario, low-mass haloes lose their gas, and therefore a fraction of their mass, early in the history of the Universe, stunting their growth and leading to a small ($0.1\,\mathrm{dex}$) shift in the low-mass HMF, but no overall change in slope \citep{2015MNRAS.448.2941S,2016MNRAS.457.1931S}. A precision measurement of the HMF would therefore constitute an excellent test of the predictions of the CDM model, especially at the low-mass end where the predictions of other dark matter models differ \citep[e.g. warm dark matter models;][]{2001ApJ...556...93B}.

As the HMF is not directly measurable we must instead rely on the kinematics of visible tracers orbiting in dark matter haloes, or other gravitational effects such as lensing, to provide indirect constraints. The expected structure of cold dark matter haloes implies that their circular velocity\footnote{To avoid ambiguity, we adopt a notation in which $v$ denotes velocity, and $V$, volume.} profiles $v_\mathrm{circ}$, related to the enclosed mass within radius $r$ as $v^2_\mathrm{circ}=GM(<r)/r$, have a broad plateau whose amplitude $v_\mathrm{max}$ is tightly correlated with the halo mass \citep{1996ApJ...462..563N,1997ApJ...490..493N}. The number density of sources as a function of the characteristic speed of rotation of those sources (as revealed by kinematic tracers), which we will term the `velocity function' (VF), therefore gives a means to constrain the HMF observationally. In practice, connecting the actual kinematic tracer observed -- such as a spectral line width -- and $v_\mathrm{max}$ requires some additional information and/or modelling.

In the case of the width of the $21$-cm emission line of neutral hydrogen (\HI), the maximum circular velocity of a halo in which a sufficiently extended disc of \HI\ gas is rotating is approximately half the full width at half-maximum (FWHM) of the line $w_{50}$, provided the inclination of the disc is accounted for, such that $v_\mathrm{max}\sim \frac{1}{2}w_{50}/\sin(i)$, where $i$ is the inclination angle. The need to correct for inclination is problematic, as surveys of line widths covering representative volumes currently do not resolve the spatial structure of the gas, necessitating reliance on optical imaging to estimate $i$ \citep[e.g.][]{2010MNRAS.403.1969Z}. There are numerous additional issues which need to be accounted for in an attempt to obtain $v_\mathrm{max}$ from a measurement of $w_{50}$ \citep[see e.g.][for detailed discussions, in addition to the references in the following list]{2001ApJ...563..694V,2017AnA...607A..13V}: the gas orbits may not be circular \citep{2016MNRAS.463L..69M}; the gas disc may not be sufficiently extended to reach the flat portion of the circular velocity curve \citep{2015MNRAS.453.2133B,2016MNRAS.463.4052P,2016MNRAS.455.3841B,2016MNRAS.463L..69M,2017ApJ...850...97B}; the gas disc may not lie in a single plane \citep{2016MNRAS.463.4052P}; the system may not be in dynamical equilibrium; the emission may be confused with neighbouring sources \citep{2015MNRAS.449.1856J,2019MNRAS.488.5898C}; the gas disc may be partially supported by turbulent or thermal pressure \citep{2015MNRAS.453.2133B,2016MNRAS.463.4052P}; etc. There have been several attempts to work around these issues and recover the HMF of gas-rich galaxies (e.g. \citealp{2009ApJ...700.1779Z,2010MNRAS.403.1969Z}; \citealp{2015AnA...574A.113P} -- hereafter \citetalias{2015AnA...574A.113P}; \citealp{2019ApJ...886L..11L,2021arXiv210803253D}). While in principle each of the issues may be addressed directly with sufficiently detailed observations and modelling, assembling a large enough sample of objects to pursue this approach has so far proven to be prohibitively expensive.

An observationally more straightforward approach is to simply measure the number density of sources as a function of $w_{50}$: the \HI\ velocity width function (\HIWF). Using the \HIWF\ as a constraint on cosmology then requires a prediction for the \HIWF\ expected in a given model, and some understanding of the possible degeneracies in this prediction across various models, but this is in principle easier to achieve than it is to solve the inverse problem of inferring the HMF from the \HIWF\ \citep{2011ApJ...739...38P}. These considerations lead us to focus on the \HIWF, and omit further discussion of the \HIVF, in this work.

The first direct measurements of the \HIWF\ were by \citet{2009ApJ...700.1779Z} and \citet{2010MNRAS.403.1969Z}, using an early partial ($6$~per~cent) release of the Arecibo Legacy Fast ALFA\footnote{Arecibo L-band Feed Array.} ({\AA}) survey, and the \HI\ Parkes All-Sky Survey ({\HI}\,PASS), respectively. It was immediately recognized (and had been anticipated, from indirect estimates) that the observed low-velocity width slope would be very difficult to reconcile with theoretical expectations -- there is an apparently severe overabundance of haloes predicted to exist which, apparently, fail to host observable galaxies. This result has been confirmed repeatedly (\citealp{2011ApJ...739...38P,2015MNRAS.454.1798K}; \citetalias{2015AnA...574A.113P}), including by the measurement we present below, and possible means to reconcile theory and measurement further discussed in the literature \citep[e.g.][]{2009ApJ...700.1779Z,2015MNRAS.453.2133B,2015MNRAS.450.3920B,2016MNRAS.463L..69M,2016AnA...591A..58P,2017ApJ...850...97B,2019MNRAS.482.5606D}. We defer further discussion of this fascinating topic to future work. In the present study, we present a new measurement of the \HIWF\ based on the catalogue from the now-completed \AA\ survey. We focus on testing the compatibility of the data with a hypothesis that is well-motivated in the {$\Lambda$}CDM cosmology: that the HMF, as encoded in the \HIWF, is spatially invariant. That is, different sub-volumes of the full survey should have similar HMFs, which may be reflected by a similarity between their {\HIWF}s.

Spatial variations in the \HIWF\ have been measured before. \citet{2009ApJ...700.1779Z} were able to measure a factor of $\sim 3$ difference in number density in between source counts in a high and a low density region in a very early \AA\ survey catalogue, despite having only $15$ detected sources in the lower density region. In the context of estimating a systematic uncertainty on their (total) measurement, \citet[][see their fig.~7]{2010MNRAS.403.1969Z} noted that while the overall number density in four quadrants of {\HI}\,PASS differed noticeably, the overall shape of the \HIWF\ appeared similar in each. A similar situation is hinted at in fig.~8 of \citet{2011ApJ...739...38P}, showing the \HIWF\ for two regions on the sky covered by an early portion of the \AA\ survey, but these authors comment on this point only very briefly. \citet{2014MNRAS.444.3559M} compared the \HIWF\ of \AA\ galaxies in voids with those in the walls of the cosmic web. Their analysis suggests that the two populations are inconsistent with being drawn from a single underlying distribution, but they conclude that statistical uncertainties prevent them from claiming a significant difference in the \HIWF\ shape. Below, we explore spatial variations in the \HIWF\ leveraging the higher precision afforded by the large number ($\sim 2\times10^4$) of extragalactic sources detected in the full \AA\ survey.

This article is structured as follows. We outline the characteristics of the \AA\ survey and catalogue, and our selection of galaxies therefrom, in Sec.~\ref{SecData}. Our methodology for the measurement of the \HIWF, using the $1/V_\mathrm{eff}$ estimator, is described in Sec.~\ref{SecMethod}. We present our measurement, including separately for independent subvolumes of the survey, in Sec.~\ref{SecResults}, and discuss its implications in Sec.~\ref{SecDiscussion}. We summarise in Sec.~\ref{SecConc}

\section{The \AA\ survey}
\label{SecData}

The \AA\ survey \citep{2005AJ....130.2598G} mapped $\sim 7000\,\mathrm{deg}^2$ of sky at $21$-cm wavelengths out to distances of $\sim 250\,\mathrm{Mpc}$ ($cz\lesssim 18000\,\mathrm{km}\,\mathrm{s}^{-1}$). The survey is composed of two contiguous areas on sky, one in the northern Galactic hemisphere, visible from Arecibo at night during the spring, the other in the southern Galactic hemisphere, visible in the autumn. Following \citet[][hereafter \citetalias{2018MNRAS.477....2J}]{2018MNRAS.477....2J}, we will refer to the two areas as the `spring' and `fall' regions, respectively\footnote{See their fig.~1 and tables D1--D4 for details of the survey footprint; we use the fiducial, not the `strict' footprint throughout this work.}.

Candidate extragalactic \HI\ sources were identified using a matched-filtering approach \citep{2007AJ....133.2087S}, supplemented by some sources identified by direct inspection of the raw data cubes. This initial catalogue was curated by hand to confirm or reject each individual detection, and to assign optical counterparts to detections wherever possible, resulting in the $\alpha.100$ extragalactic source catalogue described in \citet{2018ApJ...861...49H}. The catalogue lists the coordinates (for the \HI\ and associated optical source), redshifts, $21$-cm line flux density and width, distance, signal-to-noise ratio and \HI\ mass of sources, and their uncertainties where relevant; we refer to \citet{2018ApJ...861...49H} for details of the determination of these parameters. The `$100$' of $\alpha.100$ refers to the 100~per~cent completed survey. To compare with earlier work, we also make some use of the earlier $\alpha.40$ catalogue \citep[][hereafter \citetalias{2011AJ....142..170H}]{2011AJ....142..170H}, which is very similar but covers only 40~per~cent of the total survey area on sky.

We define a selection of sources from the $\alpha.100$ catalogue for use in measuring the \HIWF, closely following the equivalent selection used in \citetalias{2018MNRAS.477....2J} for the \HI\ mass function (\HIMF). We use all `Code 1' (i.e. $\mathrm{S}/\mathrm{N}>6.5$) sources whose \HI\ coordinates fall within the survey footprint \citepalias[][tables D1--D4]{2018MNRAS.477....2J} and have a recessional velocity in the CMB frame of $<15000\,\mathrm{km}\,\mathrm{s}^{-1}$. We adopt the same minimum line width at FWHM\footnote{We use $\log$ and $\log_{10}$ to denote the natural and base-10 logarithms, respectively.} $\log_{10}(w_{50}/\mathrm{km}\,\mathrm{s}^{-1})>1.2$ as \citetalias{2018MNRAS.477....2J}, but \emph{do not} impose a minimum \HI\ mass (\citetalias{2018MNRAS.477....2J} used $\log_{10}(M_\mathrm{HI}/\mathrm{M}_\odot)>6$). This includes $4$ additional sources. Unlike the low-mass end of the \HIMF, the low-velocity end of the \HIWF\ is somewhat sensitive to this choice (though the counting uncertainties are, of course, large). We use only sources above the bivariate 50~per~cent completeness of the survey, which depends on flux and line width. Rather than the limit derived from the $\alpha.40$ data reported in \citetalias{2011AJ....142..170H}, we use an updated completeness limit (Eq.~\ref{EqComp}) derived from the $\alpha.100$ data, see Appendix~\ref{AppComp} for details. These cuts yield a sample of $21827$ sources, of which $13620$ and $8207$ are in the spring and fall regions, respectively -- the bulk of the difference with respect to the sample of \citetalias{2018MNRAS.477....2J} is due to the updated completeness limit.

Our initial measurement of the $\alpha.100$ \HIWF\ included one obvious outlier point: the estimate in a bin centred at $\sim 90\,\mathrm{km}\,\mathrm{s}^{-1}$ was offset upward by a factor of $\sim 2$ relative to adjacent bins, and had a statistical uncertainty a factor of $\sim 6$ greater than those of its neighbours. We determined that this irregularity was driven primarily by two sources (AGC~749235, AGC~220210). Both of these are in the vicinity on the sky of the `Coma I cloud', a group of galaxies at a distance of $\sim 16.1\,\mathrm{Mpc}$ and with an unusually large peculiar velocity of $-840\,\mathrm{km}\,\mathrm{s}^{-1}$ \citep[e.g.][see their fig.~10]{2018AstBu..73..124K}. AGC~749235 has a distance of $D=6.2\pm1.3\,\mathrm{Mpc}$ in the $\alpha.100$ catalogue, but the appearance of its optical counterpart PGC~5059199 suggests a larger distance \citep{2019AstBu..74....1K}. We have been unable to locate a redshift-independent distance for AGC~220210, listed at $2.8\pm0.6\,\mathrm{Mpc}$ in $\alpha.100$, or its optical counterpart SDSS~J121323.34+295518.3, in the literature. However, this is clearly the same object as KK~127, at 121322.7+295518 (J2000), which has as Tully-Fisher distance of $17.3\,\mathrm{Mpc}$ \citep[][table~3]{2018AstBu..73..124K}. We have adopted distances of $16.1\pm4.3$ and $17.3\pm3.5\,\mathrm{Mpc}$ for AGC~749235 and AGC~220210, respectively, and have re-derived their \HI\ masses accordingly. There are likely other \AA\ sources with distance errors (beyond those reflected in the uncertainties quoted in the $\alpha.100$ catalogue) -- these remain a source of systematic uncertainty for the \HIWF\ measurement which we will not attempt to address further in the present study. We note that, following \citetalias{2018MNRAS.477....2J} (but unlike \citetalias{2015AnA...574A.113P}), we prefer to retain nearby galaxies in our analysis in order to prevent a possible artificial suppression of the low-velocity end of the \HIWF.

\section{The $1/V_\mathrm{\lowercase{eff}}$ maximum likelihood estimator}
\label{SecMethod}

The $1/V_\mathrm{eff}$ maximum likelihood estimator \citep[as described in][sec.~2, including discussion of the difference with respect to the 2DSWML estimator]{2005MNRAS.359L..30Z} is very similar to, but subtly distinct from, the bivariate step-wise maximum likelihood (2DSWML) estimator \citep[e.g. as described in][appendix~B]{2010ApJ...723.1359M}. We refer to the above references for full details, but, briefly, the $1/V_\mathrm{eff}$ method determines the effective volume in which each source observed in the course of a survey would have been detectable by that survey, accounting for the large-scale structure in the actual surveyed volume\footnote{The values $1/V_\mathrm{eff}$ are loosely analogous to the $1/V_\mathrm{max}$ weights in the classical estimator of \citet{1968ApJ...151..393S}.}. The estimator is two-dimensional in the sense that the completeness limit of the \AA\ survey is a function of both the integrated $21$-cm line flux $S_{21}$, and the line width $w_{50}$. We adopt the $50$~per~cent completeness limit\footnote{This definition approximates the sensitivity function of the survey as a step function, transitioning from a value of $0$ to a value of $1$ at the 50~per~cent completeness limit. The true transition is a smooth gradient, but we adopt this approximation to facilitate numerical calculations, following \citetalias{2015AnA...574A.113P}.} given in \citetalias{2011AJ....142..170H} (eqs.~4 \& 5). By summing the `weights' $1/V_\mathrm{eff}$ of sources in 2D bins in \HI\ mass $M_\mathrm{HI}$, determined as usual from the flux and distance as
\begin{equation}
  M_\mathrm{HI}/\mathrm{M}_\odot=2.36\times 10^5 (D/\mathrm{Mpc})^2(S_{21}/\mathrm{Jy}\,\mathrm{km}\,\mathrm{s}^{-1}),\label{eq-flux}
\end{equation}
and $w_{50}$, one obtains the \HI\ mass-velocity width function. Summing this along the velocity-width axis yields the \HIMF, while summation along the mass axis gives the \HIWF. The \HIMF\ and \HIWF\ are therefore `marginalisations' along the two axes of the same underlying distribution.

\begin{figure*}
  \includegraphics[width=\columnwidth]{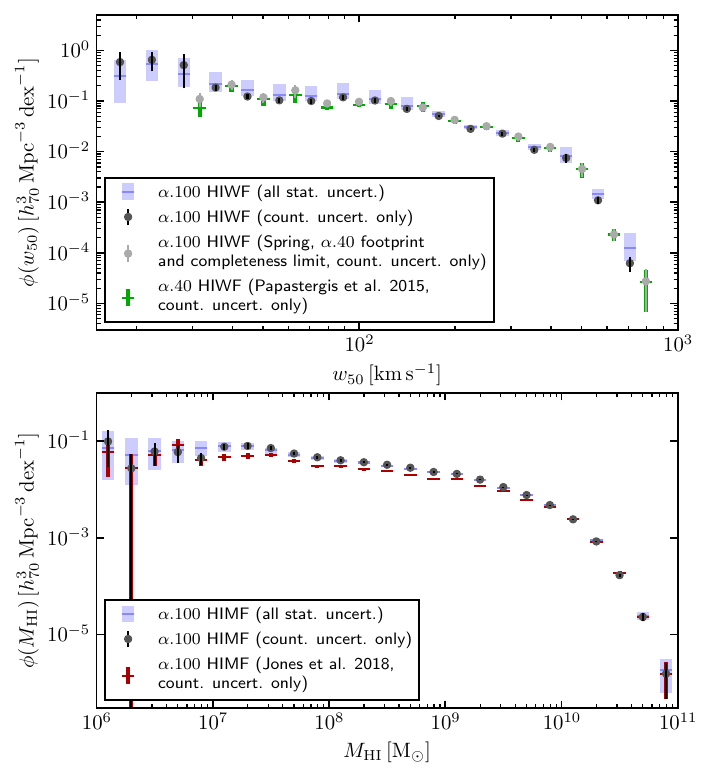}
  \includegraphics[width=\columnwidth]{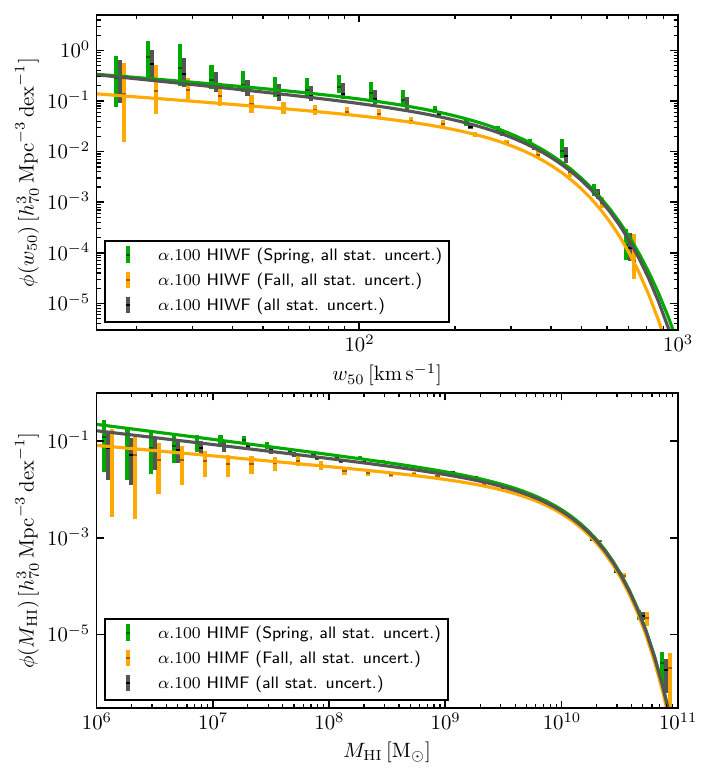}
  \caption{\emph{Upper left:} The \HIWF\ as measured from the complete $\alpha.100$ release of the \AA\ survey. The black points with error bars show the measurement accounting only counting (Poisson) uncertainties, while the blue bars and boxes also take into account distance, flux and line-width measurement uncertainties. All error bars represent 68~per~cent confidence intervals. We show for comparison the measurement of \citetalias{2015AnA...574A.113P} (green crosses) based on the $\alpha.40$ catalogue, and our measurement if we use the $\alpha.100$ catalogue cut back to the same area on sky as used by \citetalias{2015AnA...574A.113P}, and reverting to the \citetalias{2011AJ....142..170H} completeness limit (grey points). \emph{Lower left:} Our measurement of the \HIMF\ measured from the $\alpha.100$ catalogue accounting only for counting uncertainties (black points), and for all statistical uncertainties (blue bars and boxes), which closely reproduces the previously published measurement by \citetalias{2018MNRAS.477....2J} (red crosses). \emph{Upper right:} The \HIWF\ measured from the $\alpha.100$ catalogue, separately for the \AA\ spring (green) and fall (orange) sky areas; uncertainties are 68~per~cent confidence intervals and include all statistical uncertainties. The black bars and boxes are identical to the blue ones from the upper left panel; the green and orange ones have been slightly offset horizontally for clarity. The curves show the modified Schechter functions (Eq.~\ref{eq:modschechter}) describing the data (see Sec.~\ref{SecResults}). The shapes of the orange and green curves are consistent: within the uncertainties, they differ only in their normalisations. \emph{Lower right:} The \HIMF\ measured from the $\alpha.100$ catalogue, separately for the spring (green) and fall (orange) sky areas -- the black bars and boxes reproduce the blue ones from the lower left panel. The curves show the Schechter functions (Eq.~\ref{eq:schechter}) describing the data (see Sec.~\ref{SecResults}). The shapes of the curves differ significantly, as also reported by \citetalias{2018MNRAS.477....2J} (see their figs.~3 and 7).\label{fig:hiwf}}
\end{figure*}

We use the same implementation of the $1/V_\mathrm{eff}$ estimator as was used by \citetalias{2015AnA...574A.113P}. We have verified that we can exactly reproduce their measurement of the \HIWF\ (their fig.~2, reproduced in the upper left panel of Fig.~\ref{fig:hiwf} with green cross symbols) by reverting to the $\alpha.40$ catalogue and completeness limit as input, and approximately by trimming the $\alpha.100$ catalogue to the spring $\alpha.40$ footprint, using the $\alpha.100$ completeness limit (shown with grey points in the upper left panel of Fig.~\ref{fig:hiwf}). As a further check of consistency with previously published results, we have verified that we reproduce the \HIMF\ of \citetalias{2018MNRAS.477....2J} (their fig.~2, reproduced with red crosses in the lower left panel of Fig.~\ref{fig:hiwf}). Our determination based on the $\alpha.100$ catalogue, shown with the black points, lies systematically above this previous measurement. However, if we revert to the $\alpha.40$ completeness limit (\citetalias{2011AJ....142..170H}, eqs.~4 \& 5), our measurement closely follows and is fully statistically consistent with this prior measurement. A detailed explanation of the few remaining small differences with respect to these prior results is given in Appendix~\ref{AppPreviousResults}.

\section{The \AA\ \HIWF}
\label{SecResults}

Our measurement of the \HIWF\ based on the full $\alpha.100$ catalogue is shown with the black points and error bars in the upper left panel of Fig.~\ref{fig:hiwf}. In this case we account only for counting (Poisson) uncertainties; other statistical uncertainties, and systematic uncertainties, are discussed below. This measurement appears to be in reasonable agreement with the determination by \citetalias{2015AnA...574A.113P} (green crosses), but this is due to a chance partial cancellation of two competing systematic effects. First, that sample is drawn only from the spring region of the survey \citepalias[the $\alpha.40$ subset of the $\alpha.100$ spring region, see][fig.~1]{2011AJ....142..170H} -- if we restrict our galaxy sample to same region on sky, we obtain a measurement with a higher overall normalisation. However, by chance, our updated completeness limit (see Appendix~\ref{AppComp}) also tends to increase the normalisation of the \HIWF -- were the \citetalias{2015AnA...574A.113P} repeated with a matching completeness limit, it would lie significantly above our $\alpha.100$ measurement (see also Appendix~\ref{AppPreviousResults}). In summary, our measurement and that of \citetalias{2015AnA...574A.113P} should be taken to differ significantly, but the reason for this is clear: the regions of the sky covered by the two input catalogues have significantly different overall number densities of \HI\ sources.

Although many previous analyses report only the counting uncertainties on the \HIWF, the measurement uncertainties in distance, flux and velocity-width for each galaxy also make a significant contribution to the statistical uncertainty budget. We estimate the total statistical uncertainty by randomly drawing a new value for each of $D$, $S_{21}$ and $w_{50}$ from a Gaussian distribution with centre and width given by the values and uncertainties reported in the $\alpha.100$ catalogue, rejecting any unphysical (e.g. negative) values, then re-derive all derived quantities (e.g. \HI\ mass). We then measure the \HIWF\ for the resulting catalogue, again using the $1/V_\mathrm{eff}$ estimator. We repeat this process $1000$ times and average the probability distributions in each velocity-width bin, then adopt the median and $16^\mathrm{th}$ to $84^\mathrm{th}$ percentile interval as an estimate of the total statistical uncertainty. This is shown with the blue bars and boxes in the upper left panel of Fig.~\ref{fig:hiwf}. (We repeat the same exercise for the \HIMF, the result is shown with the blue bars and boxes in the lower left panel of Fig.~\ref{fig:hiwf}.)

These blue bars and boxes are reproduced in black in the upper right panel of Fig.~\ref{fig:hiwf}. We have also repeated the same process for two independent subsets of our $\alpha.100$ sample, corresponding to the spring (green bars and boxes) and fall (orange bars and boxes) areas on sky. The two subsets of the survey are treated fully independently, i.e. we re-determine the $V_\mathrm{eff}$ values for each source using only the sources in the same subset in the calculation. The spring and fall {\HIWF}s have similar shapes (quantified below), but are offset `vertically' in overall number density. This further illustrates the main driver of the disagreement with the $\alpha.40$-based measurement of \citetalias{2015AnA...574A.113P}: that sample is drawn only from the spring region, which lies slightly above the full $\alpha.100$ \HIWF.

\begin{table*}
  \caption{Parameter values and 95~per~cent confidence intervals for the (modified) Schechter functions describing the \HIMF\ and \HIWF. We give both the counting (Poisson) uncertainties alone, and the total statistical uncertainties. See Sec.~\ref{SecDiscussion} for a discussion of the uncertainty budget of the \HIWF, including systematics, and see \citetalias{2018MNRAS.477....2J} for a detailed discussion of the uncertainty budget of the $\alpha.100$ \HIMF. The values given here are for the $2.5^\mathrm{th}$, $50^\mathrm{th}$ and $97.5^\mathrm{th}$ percentiles of the marginalized distributions; the curves in Fig.~\ref{fig:hiwf} correspond to the median values when all statistical uncertainties are accounted for -- see Sec.~\ref{SecResults} for further details.\label{tab:fit}}
  \begin{tabular}{lcccc}
    \hline
    \multicolumn{5}{c}{Schechter function parameters (\HIMF)}\\
    & $\log_{10}\phi_\star$ & $\log_{10}M_\star$ & $\alpha$ & \\
    Sample & $[h_{70}^3\,\mathrm{Mpc}^{-3}\,\mathrm{dex}^{-1}]$ & $[\mathrm{M}_\odot]$ & & \\
    \hline
    \multicolumn{5}{c}{Counting (Poisson) uncertainties only}\\
    \hline
    $\alpha.100$ & $-2.23^{+0.01}_{-0.01}$ & $9.90^{+0.01}_{-0.01}$ & $-1.27^{+0.01}_{-0.01}$ &\\
    $\alpha.100$ Spring & $-2.18^{+0.02}_{-0.02}$ & $9.89^{+0.01}_{-0.01}$ & $-1.27^{+0.01}_{-0.01}$ & \\
    $\alpha.100$ Fall & $-2.29^{+0.02}_{-0.02}$ & $9.89^{+0.02}_{-0.02}$ & $-1.22^{+0.02}_{-0.02}$ & \\
    \hline
    \multicolumn{5}{c}{All statistical uncertainties}\\
    \hline
    $\alpha.100$ & $-2.26^{+0.02}_{-0.02}$ & $9.92^{+0.01}_{-0.01}$ & $-1.29^{+0.02}_{-0.02}$ &\\
    $\alpha.100$ Spring & $-2.23^{+0.03}_{-0.02}$ & $9.92^{+0.02}_{-0.02}$ & $-1.31^{+0.03}_{-0.02}$ &\\
    $\alpha.100$ Fall & $-2.30^{+0.03}_{-0.03}$ & $9.90^{+0.02}_{-0.02}$ & $-1.22^{+0.03}_{-0.03}$ &\\
    \hline
    \hline
    \multicolumn{5}{c}{Modified Schechter function parameters (\HIWF)}\\
    & $\log_{10}\phi_\star$ & $w_\star$ & $\alpha$ & $\beta$\\
    Sample & $[h_{70}^3\,\mathrm{Mpc}^{-3}\,\mathrm{dex}^{-1}]$ & $[\mathrm{km}\,\mathrm{s}^{-1}]$ & & \\
    \hline
    \multicolumn{5}{c}{Counting (Poisson) uncertainties only}\\
    \hline
    $\alpha.100$ & $-1.67^{+0.21}_{-0.16}$ & $300^{+48}_{-58}$ & $-0.56^{+0.21}_{-0.16}$ & $2.1^{+0.4}_{-0.3}$ \\
    $\alpha.100$ Spring & $-1.51^{+0.28}_{-0.22}$ & $279^{+65}_{-78}$ & $-0.50^{+0.31}_{-0.22}$ & $1.9^{+0.5}_{-0.4}$ \\
    $\alpha.100$ Fall & $-2.14^{+0.15}_{-0.12}$ & $382^{+37}_{-45}$ & $-0.82^{+0.16}_{-0.13}$ & $2.7^{+0.5}_{-0.4}$ \\
    \hline
    \multicolumn{5}{c}{All statistical uncertainties}\\
    \hline
    $\alpha.100$ & $-1.67^{+0.76}_{-0.42}$ & $307^{+127}_{-226}$ & $-0.63^{+1.12}_{-0.41}$ & $2.0^{+1.3}_{-1.0}$ \\
    $\alpha.100$ Spring & $-1.46^{+0.74}_{-0.58}$ & $268^{+173}_{-243}$ & $-0.50^{+1.75}_{-0.56}$ & $1.8^{+1.6}_{-1.0}$ \\
    $\alpha.100$ Fall & $-1.85^{+0.64}_{-0.36}$ & $310^{+105}_{-251}$ & $-0.48^{+1.64}_{-0.40}$ & $2.1^{+1.1}_{-1.2}$ \\
    \hline
  \end{tabular}
\end{table*}

The \HI\ \emph{mass} function has a significantly different shape in the two sky regions: the spring sky has a steeper low-mass slope, and a slightly higher `knee' mass, where the approximately exponential decline at high masses begins. The difference in shape was extensively discussed in \citetalias{2018MNRAS.477....2J}, to which we refer for full details. We illustrate this difference in shape in the lower right panel of Fig.~\ref{fig:hiwf}, which is analogous to the upper right panel, but for the \HIMF, rather than the \HIWF. We parametrize the shape of the \HIMF\ with a Schechter function $\phi_\mathrm{S}$ \citepalias[e.g.][]{2018MNRAS.477....2J}:
\begin{equation}
  \frac{\phi_\mathrm{S}(M_\mathrm{HI})}{\phi_\star} = \log(10)\left(\frac{M_\mathrm{HI}}{M_\star}\right)^{\alpha+1} \mathrm{e}^{-M_\mathrm{HI}/M_\star}\label{eq:schechter}
\end{equation}
where $\phi_\star$ is the overall normalisation, $M_\star$ is the `knee mass' and $\alpha+1$ is the low-mass slope. We constrain the values of these parameters by sampling the posterior probability distribution for the parameters for each of our $1000$ re-sampled catalogues using a Markov chain Monte Carlo (MCMC) sampling\footnote{We use the affine-invariant ensemble sampler implementation \texttt{emcee} \citep{2013PASP..125..306F}.}. We use the likelihood function $\log\mathcal{L}\propto-\sum_i ((\phi_i-\phi_\mathrm{S}(M_{\mathrm{HI},i}))/\sigma_i)^2$ where the index $i$ runs over the points, and $\sigma_i$ is the Poisson error estimate on $\phi_i$. We then take the union of the $1000$ samplings as an estimate of the posterior probability distribution accounting for all statistical uncertainties\footnote{We could also have chosen to use the \HIMF\ with our estimates of the full statistical uncertainties, i.e. the bars and boxes in the right panels of Fig.~\ref{fig:hiwf}, to constrain the Schechter function parameters directly, but we feel that our adopted approach likely better reflects the true confidence intervals for the parameters. However, we have also checked explicitly that the two approaches give statistically consistent results, for both the \HIMF\ and \HIWF.}. The median parameter values and their 95~per~cent confidence intervals are summarized in Table~\ref{tab:fit}, and these median parameter values are used to plot the curves in the right panels of Fig.~\ref{fig:hiwf}.

\begin{figure*}
  \includegraphics[width=\textwidth]{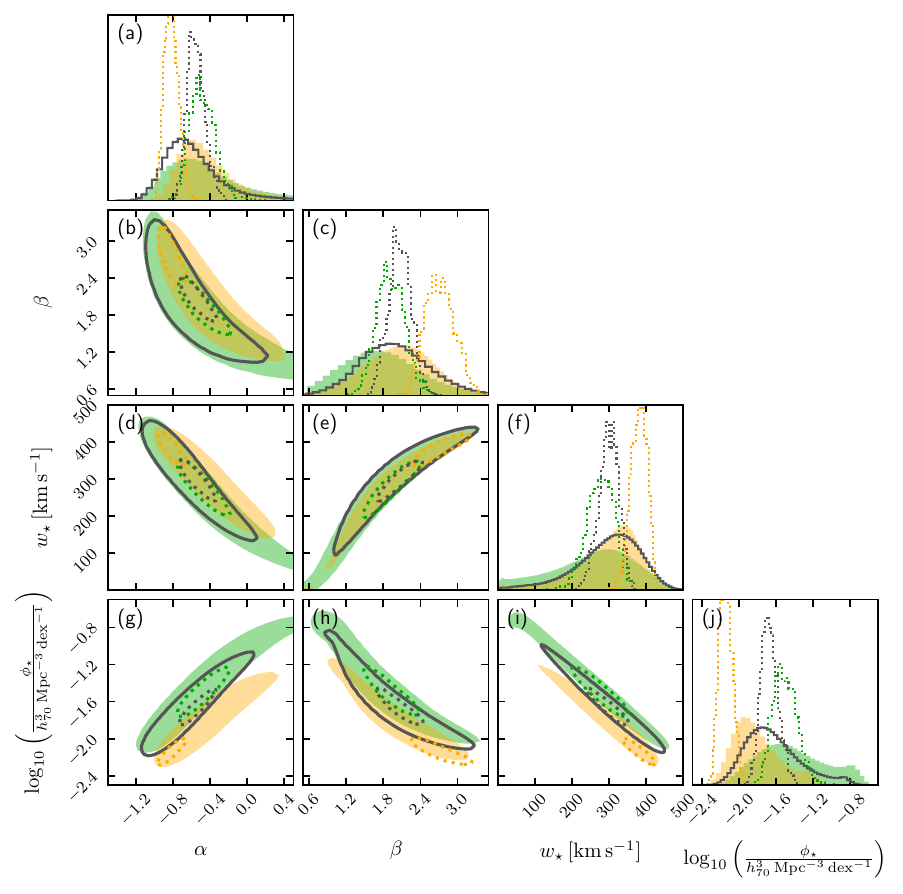}
  \caption{One- and two-dimensional marginalised posterior probability distributions for the parameters of the modified Schechter function (Eq.~\ref{eq:modschechter}). All contours are drawn at the $95$~per~cent confidence level. Dark grey dotted contours and histograms correspond to the fit to the \HIWF\ of the full $\alpha.100$ sample when only counting (Poisson) uncertainties are accounted for. Dark grey solid contours and histograms are similar, but include all statistical uncertainties (see Sec.~\ref{SecResults} for details). Dark green and orange contours and dotted histograms show the parameter constraints for the spring and fall sky areas, respectively, accounting only for counting uncertainties, while the light green and orange contours and filled histograms show the constraints when all statistical uncertainties are accounted for. The confidence regions for the spring and fall regions overlap significantly, except where the normalisation, $\phi_\star$, is involved (panels g, h, i, j). A similar figure for the \HIMF\ parameters is included in Appendix~\ref{AppMFCorner}.\label{fig:corner}}
\end{figure*}

The situation is qualitatively different when we compare the \HI\ \emph{width} function for the spring and fall skies. We quantify their shapes, proceeding similarly to above, but using a modified Schechter function $\phi_\mathrm{MS}$, as is conventional for the \HIWF\ \citepalias[e.g.][]{2015AnA...574A.113P}:
\begin{equation}
  \frac{\phi_\mathrm{MS}(w_{50})}{\phi_\star} = \log(10)\left(\frac{w_{50}}{w_\star}\right)^\alpha\mathrm{e}^{-\left(w_{50}/w_\star\right)^\beta}\label{eq:modschechter}
\end{equation}
where $\phi_\star$ is again the normalisation, $w_\star$ is the `knee velocity width', $\alpha$ is the low-mass slope\footnote{Note this is not $\alpha+1$ as in Eq.~\ref{eq:schechter}; we have defined the functions for ease of comparison with earlier work based on \AA.}, and the additional parameter $\beta$ controls how sharply the exponential decay enters at high velocity widths. The parameter values are again summarized in Table~\ref{tab:fit}, and we show the one- and two-dimensional marginalised posterior probability distributions in Fig.~\ref{fig:corner} (dark grey solid, and green and orange filled, contours and histograms). While there is a significant difference in the overall normalisation $\phi_\star$ (panels g, h, i and j in Fig.~\ref{fig:corner}), the three parameters $(\alpha, \beta, w_\star)$ describing the shape of the \HIWF\ in the spring and fall skies have statistically consistent values. This implies that, unlike the \HIMF, the \HIWF\ of gas-rich galaxies is consistent with having a universal shape, as might arise from an underlying universal HMF. However, we note that the confidence intervals are noticeably wider than for the cognate parameters describing the shape of the \HIMF. We next discuss possible caveats to, and implications of, this observation.

(Our measured \HIMF\ and \HIWF\ amplitudes and uncertainties, including all those shown in Fig.~\ref{fig:hiwf} (for $\alpha.100$), are tabulated in Appendix~\ref{AppTables}.)

\section{Discussion}
\label{SecDiscussion}

We first briefly comment further on the statistical uncertainties in our measurements (Sec.~\ref{SubsecStat}) before moving on to possible systematic uncertainties (Sec.~\ref{SubsecSys}). Finally, we outline our interpretation of our measurement in Sec.~\ref{SubsecInterpret}.

\subsection{Statistical uncertainties}
\label{SubsecStat}

When accounting only for counting uncertainties, the 95~per~cent confidence regions for the spring and fall regions involving the normalisation $\phi_\star$ (Fig.~\ref{fig:corner} panels g, h and i, dotted green and orange contours) are well-separated. When all statistical uncertainties are accounted for, although the one-dimensional marginalised distributions for $\phi_\star$ (panel j, filled histograms) overlap much more than they did when accounting only for counting uncertainties (green and orange dotted histograms), it is clear from some of the two-dimensional projections (especially panels g and i) that there is still evidence for a difference in the \HIWF\ normalisation.

We note that the importance of the measurement uncertainties relative to the counting uncertainties is much greater for the \HIWF\ than it is for the \HIMF\ \citepalias[e.g.][table~1]{2018MNRAS.477....2J} -- this is because the $w_{50}$ uncertainties have a negligible influence on the \HIMF, but contribute significantly to the \HIWF\ uncertainty budget.

By happenstance, the confidence region for the fall $\alpha.100$ sample when not accounting for measurement uncertainties is located in a low probability density region, near the edge\footnote{Though not shown, in all cases the peaks of all two-dimensional marginalised distribution in Fig.~\ref{fig:corner} lie approximately at the centres of the plotted $95$~per~cent confidence regions.} of its parent distribution (i.e. including all measurement uncertainties), as can be seen in Fig.~\ref{fig:corner} (panels b, d, e, g, h, i). We point out this statistical curiosity only to note that it does not influence our interpretation below.

We include a figure similar to Fig.~\ref{fig:corner} for the \HIMF\ parameters in Appendix~\ref{AppMFCorner}, Fig.~\ref{fig:corner_mf}.

\subsection{Systematic uncertainties}
\label{SubsecSys}

In the context of our main focus in this work, that is, assessing whether the shapes of the {\HIWF}s in the spring and fall sky regions are indeed indistinguishable at a level of precision accessible using $\alpha.100$, we are primarily interested in possible systematic errors which could have independent effects in the two regions. Systematic errors expected to affect parameter estimates in both regions in the same way will widen the confidence intervals somewhat, but not affect our qualitative conclusions. We refer to \citetalias[][e.g. table~1]{2018MNRAS.477....2J} as a guide to some of the sources of systematic uncertainty to be considered.

In addition, we discuss in Sec.~\ref{SubsubsecUnseen} a systematic effect which has not to our knowledge been previously explored in detail and may see the low-velocity slope of the \HIWF\ increase by $\sim 40$~per~cent in future, deeper surveys.

\subsubsection{Survey boundary}

The measurement of the \HIWF\ is sensitive to the survey boundary chosen. The fiducial boundary which we have used throughout our analysis aims to maximise the area while avoiding the irregular coverage near the edges due to slight differences in the beginning and end in right ascenscion of \AA\ drift scans. We have explicitly checked the effect of instead adopting the `strict' boundary defined in \citetalias{2018MNRAS.477....2J} tables~D1--D4: this slightly widens the confidence intervals for all parameters, but leaves our conclusions unaffected.

\subsubsection{Sample variance}

We have already established that the \HIWF\ is consistent with having the same shape in the spring and fall areas of the \AA\ survey, within the statistical uncertainties. We now explore the possibility of spatial variations on smaller scales. We follow \citetalias{2018MNRAS.477....2J} and quantify this by jackknifing the calculation of the \HIWF: we split the survey area into $42$ approximately equal-area sub-regions ($27$ and $15$ in the spring and fall regions, respectively) and determine the \HIWF, removing each sub-region in turn. For the spring region this yields an estimate of the systematic uncertainty due to `sample variance' of $(0.006, 0.02, 2, 0.006)$ for the paramters $(\alpha, \beta, w_\star/\mathrm{km}\,\mathrm{s}^{-1}, \log_{10}(\phi_\star/h^3_{70}\,\mathrm{Mpc}^{-3}\mathrm{dex}^{-1}))$, and of $(0.02, 0.04, 5, 0.02)$ for the fall region. Even if this systematic error were to conspire to move the two posterior probability distributions directly apart in parameter space (which we deem unlikely), they would still overlap substantially.

\subsubsection{Distance model}

In order to estimate the influence of the distance model on our estimate, we replace the \citet{2005PhDT.........2M} flow model used in the fiducial analysis presented above with a very different model: we simply assume Hubble flow distances $D=v_\mathrm{CMB}H_0$ for all sources. We then repeat our analysis, imposing a minimum distance of $7\,\mathrm{Mpc}$ \citepalias[as in][]{2015AnA...574A.113P}, instead of $0\,\mathrm{Mpc}$, to avoid the region giving rise to the most severe distance errors due to peculiar velocities. Even with these radically different distance estimates, the posterior probability distributions (accounting for all statistical uncertainties) for $\alpha$, $\beta$ and $w_\star$ still overlap substantially.

We note that the \HIWF\ is more sensitive to distance errors than the \HIMF. This is because if, for example, the distance of a source is underestimated, its mass will consequently also be underestimated (Eq.~\ref{eq-flux}). This in turn causes the effective volume ($V_\mathrm{eff}$) in which the source would be accessible to the survey to be underestimated, causing an overestimate of the true abundance of similar sources. For the \HIWF, this leads directly to an overestimate of the number density in the bin containing the $w_{50}$ of the source. Such errors can easily be severe, as seen in Sec.~\ref{SecData} above. For the \HIMF, however, the affected bin is that of the (incorrect) mass estimated for the source, where the number density is higher than that associated with the actual source in question (due to the negative slope), leading to much smaller relative errors. An overestimated distance, on the other hand, causes $V_\mathrm{eff}$ to be overestimated, effectively down-weighting the contribution of the source to the \HIWF\ and \HIMF.

\subsubsection{Absolute flux and velocity width calibration}

The absolute flux calibration of the \AA\ survey directly influences the \HIMF\ (Schechter function) parameter $M_\star$, as adjusting the calibration directly causes a `horizontal' shift of the \HIMF. The derivation of the \HIWF, however, effectively marginalises over this effect, making it essentially insensitive to the flux calibration. The equivalent effect for the \HIWF\ would be a systematic error in the determination of $w_{50}$. The $w_{50}$ measurements in the $\alpha.100$ catalogue are corrected for instrumental broadening following \citet{2005ApJS..160..149S}; the correction depends primarily on the signal-to-noise ratio of the spectrum in question. While some residual systematic bias likely remains after this correction, it is likely to be very small, and it seems unlikely that it would affect sources in the two sky areas differently. For instance, if line widths for all low signal-to-noise sources (considering those above the cut for Code~1 sources) in the spring sky are slightly overestimated, then the same is likely to be true for all sources in the fall sky, leading to an equal `horizontal' shift in the {\HIWF}s in both regions -- we note that the distribution of signal-to-noise ratios for sources in our spring and fall \HIWF\ samples are near-identical.

\subsubsection{Adopted completeness limit}

A mismatch between the true sensitivity of the \AA\ survey and that implied by the completeness limit assumed (see Appendix~\ref{AppComp}) in the analysis in Sec.~\ref{SecResults} can lead to systematic biases in the recovered \HIMF\ and \HIWF. However, for reasonable variations in the assumed limit, the changes to both are small. For example, using the completeness limit of \citetalias{2011AJ....142..170H} derived from the $\alpha.40$ catalogue, which is offset `down' by $0.02\,\mathrm{dex}$ in $S_{21}$ relative to that we derive in the Appendix, the parameters of the modified Schechter function describing the \HIWF\ of the full survey (including all statistical errors) change by $-0.10\,h_{70}^3\,\mathrm{Mpc}^{-3}\,\mathrm{dex}^{-1}$ ($\phi_\star$), $+4\,\mathrm{km}\,\mathrm{s}^{-1}$ ($w_\star$), $+0.03$ ($\alpha$) and $0.0$ ($\beta$), in all cases by much less than our quoted uncertainties.

We have also considered whether the completeness limit of the survey may differ between the two survey regions. We assess this by following the approach of \citetalias{2011AJ....142..170H} (sec.~6), using the sources from the two survey regions as separate inputs. We find tentative evidence that the survey is slightly shallower in the fall region (offset from our fiducial $50$~per~cent completeness limit by $0.009\,\mathrm{dex}$ in $S_{21}$ at all $w_{50}$), while in the spring region the coverage may be slightly deeper (by $0.011\,\mathrm{dex}$), for a net difference between the two regions of $0.02\,\mathrm{dex}$. We note that this is one of the rare systematic uncertainties which can realistically cause opposite biases in the \HIWF\ measurements in the two regions, however the quantitative differences are small: the $0.02\,\mathrm{dex}$ offset is similar in magnitude to the offset between our fiducial $50$~per~cent completeness limit and that from \citetalias{2011AJ....142..170H} just discussed, and causes similarly small changes to the parameters describing the \HIWF\ shape. The changes in the \HIMF\ shape following reasonable changes to the assumed completeness limit are likewise small (though in this case sometimes comparable to the statistical uncertainties, which are relatively much smaller for the Schechter function parameters than the modified Schechter function parameters). Our assessment is therefore that the uncertainty in the exact completeness limit of the \AA\ survey is insufficient to qualitatively affect our conclusions.

\subsubsection{The unseen portion of the galaxy population}
\label{SubsubsecUnseen}

In the $1/V_\mathrm{eff}$ estimator, any 2D bin in $M_\mathrm{HI}$ and $w_{50}$ in which the survey has a detection count of zero makes no contribution to the \HIMF\ or \HIWF, even if the number density of such sources could be quite high and zero were detected simply because the survey is only capable of detecting them when they are very nearby. Formally, the severity of the systematic error due to this effect is unbounded: there is always space in the survey volume to hide very high abundances of low-mass and/or high-velocity-width galaxies. Its importance must therefore be evaluated in the context of some prior assumption about the properties of the intrinsic population of galaxies sampled by the survey. We show in this section that for \AA, this source of systematic error could plausibly cause the estimate of the low-velocity slope of the \HIWF\ to be too shallow by $\sim0.2$, while the impact on the \HIMF\ is likely much less severe\footnote{The two biases -- that of the low-velocity slope of the \HIWF\ and that of the low-mass slope of the \HIMF\ -- have very different magnitudes because neither the (possible) shape of the intrinsic distribution of galaxies in the $M_\mathrm{HI}$--$w_{50}$ plane, nor the selection function of the survey are close to symmetric with respect to the two relevant parameters.}. (The other parameters are also affected, but we focus on the slope to streamline our discussion.) We note that, to our knowledge, this systematic error affects nearly all\footnote{The volume-limited, optically-selected sample of \citet{2015MNRAS.454.1798K} is likely immune. Interestingly, those authors find a somewhat steeper low-width slope for the \HIWF\ than in \AA\ (e.g. their fig.~10); this discrepancy could plausibly be quantitatively resolved by the bias described in this section.} previously published measurements of both the \HIMF\ and \HIWF; the severity of the error will depend on the details of the survey and analysis in question.

\begin{figure*}
  \includegraphics[width=\textwidth]{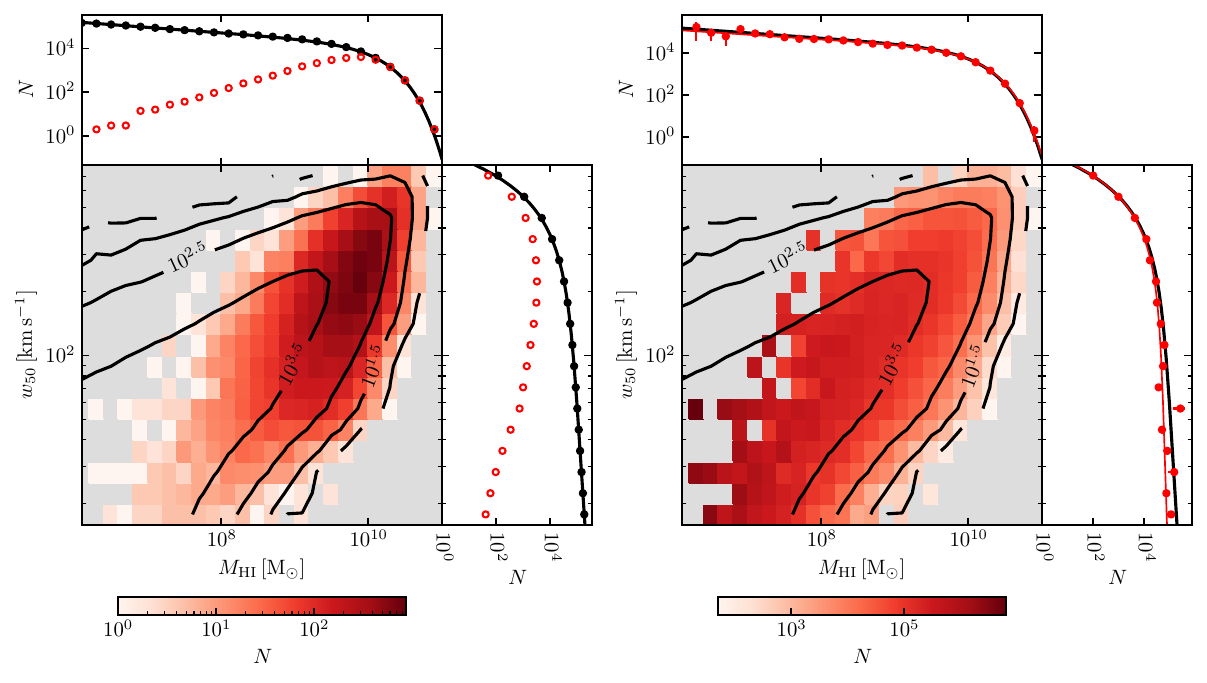}
  \caption{\emph{Left}: A simple mock survey resembling \AA\ -- mock (a) -- as described in Sec.~\ref{SubsubsecUnseen}. The contours show the `intrinsic' source population distribution as a function of $M_\mathrm{HI}$ and $w_{50}$. The sources are uniformly distributed throughout the survey volume; the heatmap shows those which have a flux and linewidth above the \AA\ 50~per~cent completeness limit, i.e. the `observed' sources. The upper inset panel shows the \HIMF\ of the intrinsic population (black points) and the observed population (red points), and the Schechter function that the mock sample is constructed to follow (black line). The right inset is similar but shows the \HIWF; the sources are drawn from the distribution corresponding to the modified Schechter function shown (black line). \emph{Right}: Similar to left, but showing the intrinsic population reconstructed from the observed population using the $1/V_\mathrm{eff}$ estimator (heatmap). The contours in the main panel and black lines in the insets are repeated from left. The insets show the recovered \HIMF\ and \HIWF\ (red points, error bars show counting errors only) and a (modified) Schechter function fit to each (red line). A significant portion of the intrinsic population is missed in the reconstruction in the region where there are no observed sources (low $M_\mathrm{HI}$, high $w_{50}$). The \HIMF\ is reasonably well-recovered despite this, but the \HIWF\ is biased toward a shallower slope at the low-velocity end.\label{fig:bias}}
\end{figure*}

To illustrate the source of the error, we construct a simple mock survey, which we label `mock~(a)'. Full details are given in Appendix~\ref{AppMocks}; we summarise here. We draw a large set of \HI\ velocity widths from a distribution described by a modified Schechter function with parameters close to those measured for \AA, $(w_\star/\mathrm{km}\,\mathrm{s}^{-1}, \alpha, \beta)=(320, -0.5, 2.2)$. Distances are randomly assigned such that the source population is uniformly distributed in the survey volume. We then assign an \HI\ mass to each source such that (i) the \HIMF\ of the population is close to that measured for \AA, within 10~per~cent at all masses of a Schechter function with parameters $(\log_{10}(M_\star/\mathrm{M}_\odot), \alpha)=(9.94, -1.25)$, and (ii) the mock source count distribution (i.e. after applying the selection function as described below) in the $M_\mathrm{HI}$--$w_{50}$ plane resembles that observed in \AA\ as closely as possible. This was achieved brute-force `by hand', for reasons detailed in the Appendix. The flux of each mock source was computed from its \HI\ mass and distance, and randomly selected sources were removed from the sample until the number with fluxes and line widths placing them above the \AA\ 50~per~cent completeness limit \citepalias[][eqs. 4 \& 5]{2011AJ....142..170H} was equal to $21827$, the total count used in our analysis in Sec.~\ref{SecResults}. We recorded two catalogues, the `intrinsic population' including all the mock sources, and the `mock survey' including only those above the completeness threshold. The $M_\mathrm{HI}$--$w_{50}$ distributions of both catalogues are shown in the main left panel of Fig.~\ref{fig:bias} with the contours and heatmap, respectively.

The right panels of Fig.~\ref{fig:bias} clearly illustrate the origin of the systematic error. The contours are repeated from the left panel -- this is the intrinsic population which the $1/V_\mathrm{eff}$ estimator should ideally reconstruct. The heatmap in this panel shows the source counts estimated for the entire survey volume using the $1/V_\mathrm{eff}$ method. While the incompleteness-corrected counts closely approximate the intrinsic population across all areas where the mock survey source count is $>0$, counts in cells where there are $0$ mock surveyed galaxies are left set to $0$. This results in an appreciable number of sources in the intrinsic population being `missed' by the estimator, predominantly toward the upper left of the figure. When the incompleteness-corrected counts are integrated along the two axes to give the \HIMF\ and \HIMF, it turns out that the input \HIMF\ is recovered accurately, $(\log_{10}(M_\star/\mathrm{M}_\odot), \alpha)=(9.94, -1.22)$, but the \HIWF\ has a slope that is noticeably too shallow at the low-velocity end, with $(w_\star/\mathrm{km}\,\mathrm{s}^{-1}, \alpha, \beta)=(321, -0.25, 2.2)$. We have repeated this exercise with many independently realised mock surveys and confirm that the effect is systematic, consistently resulting in a low-velocity slope $\alpha$ of about $-0.3$.

\begin{figure*}
  \includegraphics[width=\textwidth]{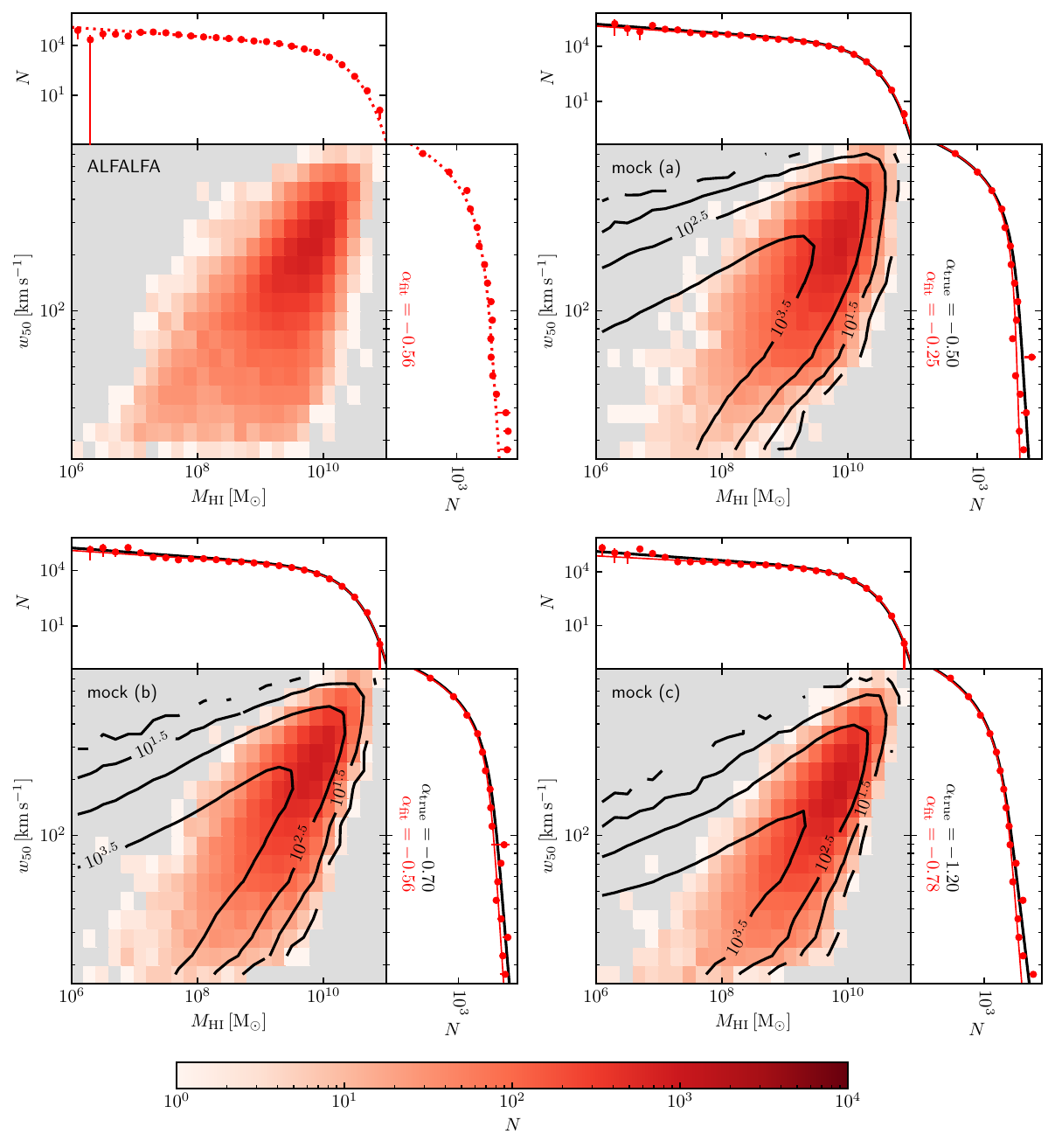}
  \caption{In the upper left group of panels, the main panel shows the observed counts in $\alpha.100$ as a function of $M_\mathrm{HI}$ and $w_{50}$. The upper and right panels show the \HIMF\ and \HIWF\ as recovered using the $1/V_\mathrm{eff}$ estimator (points with error bars), respectively, and (modified) Schechter function fits thereto (dotted lines). The main panel in the other three sets of panels show the intrinsic counts in a mock galaxy population (mocks (a), (b) and (c), as labelled -- see Sec.~\ref{SubsubsecUnseen} and Appendix~\ref{AppMocks}) with contour lines, and the mock `observed' counts for each once the \AA\ selection function is applied with the colour maps. The upper and right panels in each set show the \HIMF\ and \HIWF, respectively: the `true' function (heavy black line), the $1/V_\mathrm{eff}$ estimate (points with error bars), and a (modified) Schechter function fit to the estimate (thin red line). The intrinsic and recovered slopes of the \HIWF\ are denoted $\alpha_\mathrm{true}$ and $\alpha_\mathrm{fit}$, respectively. The slope is consistently underestimated, with the intrinsic slope of $-0.70$ for mock (b) resulting in a recovered slope of $-0.56$, matching the slope of $-0.56$ measured for $\alpha.100$. \label{fig:moremocks}}
\end{figure*}

Given that an intrinsic \HIWF\ slope of $\alpha=-0.50$ results in a slope of about $-0.3$ recovered by the $1/V_\mathrm{eff}$ estimator, it is a plausible hypothesis that a steeper intrinsic slope might result in a (biased) recovered slope close to that measured in \AA. We verify this by constructing two more mock samples following the same approach described above, but with intrinsic slopes of $\alpha=-0.7$ -- `mock~(b)' -- and $\alpha=-1.2$ -- `mock~(c)'. The mock samples are illustrated in Fig.~\ref{fig:moremocks}, similar to the left panel of Fig.~\ref{fig:bias}. These steeper intrinsic slopes are also both underestimated by the $1/V_\mathrm{eff}$ estimator, giving $\alpha=-0.56$ and $-0.78$ for mocks~(b) and (c), respectively. This confirms (i) that this bias can plausibly lead to a measurement similar to that for \AA\ ($\alpha=-0.56$) when the true slope is about $-0.7$ and (ii) that the bias likely does not cause the estimator to `saturate' (e.g. always return the same slope estimate for ever increasing intrinsic slope) over a reasonably wide range in intrinsic $\alpha$.

We make one more observation regarding the plausibility of this bias significantly affecting the \AA\ measurement. We found it challenging to reproduce the shape of the \AA\ souce count distribution in the $M_\mathrm{HI}$--$w_{50}$ plane, but had the most success when the intrinsic slope was $\alpha\sim-0.7$. This can be seen in Fig.~\ref{fig:moremocks}: the distribution for mock~(b) (lower left) is the closest match to the \AA\ distribution (upper right). Mock~(a) (upper right) has a distribution that is rather too narrow at low $w_{50}$, while mock~(c) (lower right) is somewhat too narrow everywhere. We consider this observation to be suggestive only, because the shape of the observed source count distribution depends on the shape chosen for the intrinsic distribution at fixed $w_{50}$ -- there is some freedom in this choice, and we have not exhaustively explored all possibilities.

To our knowledge, the influence of this systematic error on the \HIWF\ has not previously been thoroughly explored, though its influence on the \HIMF\ is acknowledged by \citet{2010ApJ...723.1359M}. They estimate its importance using a mock catalogue constructed by fitting the distribution of observed sources and extrapolating the tail of the distribution. We note that this is qualitatively different from our approach (Appendix~\ref{AppMocks}); we use the same form for the distribution as them but assume it describes the intrinsic population, fixing its parameters to match the observed counts once incompleteness is accounted for. Despite this, our conclusions agree: they estimate a bias of about 2~per~cent ($0.025$ out of $-1.25$) in the low-mass slope of the \HIMF; for mocks~(a), (b) and (c) the same parameter is underestimated by 2, 0 and 5~per~cent, respectively. However, their approach would likely underestimate the bias for the \HIWF.

That this bias significantly affects the low-velocity slope of the \HIWF\ shown in Fig.~\ref{fig:hiwf} is difficult to dispute: the sharp edge of the distribution toward high $w_{50}$ and low $M_\mathrm{HI}$ reconstructed by the $1/V_\mathrm{eff}$ estimator (Fig.~\ref{fig:bias} heatmap in right panel\footnote{While this is for a mock sample, the equivalent for the actual \AA\ survey is very similar, e.g. \citetalias{2015AnA...574A.113P} fig.~A.1.}) is highly unlikely to be the true edge of the galaxy population in this plane. However, the fraction of galaxies `missed' due to this error depends entirely on the shape of the distribution precisely where it is not constrained by measurement. While we judge our estimates above to be plausible, definitively settling this issue requires a deeper survey, or perhaps a search optimised for low signal-to-noise, large velocity width sources in existing surveys.

\subsection{Interpretation}
\label{SubsecInterpret}

\subsubsection{Connection to the HMF}

We have shown above that spatial variation in the \AA\ \HIWF\ is qualitatively different to that in the \HIMF. As shown in Fig.~\ref{fig:hiwf} and discussed in detail by \citetalias{2018MNRAS.477....2J}, the latter has a significantly shallower low-mass slope in the lower-density fall survey region than in the higher-density spring survey region. The \HIWF, on the other hand, has a low-velocity slope and overall shape that appear to be spatially invariant, at least within the confidence intervals implied by the data. Though there are several effects likely to cause systematic biases in the \HIWF\ measurement, we have not identified any likely to affect the two survey regions significantly differently, suggesting that the spatial invariance may not be a simple coincidence, but instead arise from an underlying spatially invariant HMF. We acknowledge that the confidence intervals for the parameters describing the the shape of the \AA\ \HIWF\ remain relatively wide and will therefore be interested to see whether the insensitivity to density on scales of a few tens of megaparsecs persists in forthcoming, larger surveys, such as WALLABY \citep{2020ApnSS.365..118K}. In this section we assume that this spatial invariance is a real feature of the galaxy population and proceed to explore its implications.

Barring some conspiratorial coincidence, the similarity in the shape of the \HIWF\ in the \AA\ spring and fall skies, despite the difference in the shape of the \HIMF, suggests a more fundamental similarity in the underlying galaxy population. We assume here the well-motivated hypothesis that the HMF in the two regions is similar except for its overall normalisation \citep[e.g.][]{2009MNRAS.399.1773C}, and explore where this assumption leads us below.

The 21-cm emission spectrum of a galaxy arises through a combination of the geometry of the system (e.g. radial surface density profile in the disc and vertical density profile falling away from the midplane; inclination) and its kinematics (rotation curve, vertical rotation velocity gradient, velocity dispersion). That the inclination-corrected line width is often a fair proxy for the maximum circular velocity of the dark matter halo comes about because most of the \HI\ gas rotates at a similar speed: most of the \HI\ is at larger radii in the disc \citep[because the central \HI\ surface density profiles of galaxies are typically fairly flat, e.g.][]{2016MNRAS.460.2143W} where the rotation curve is often flat or has a shallow gradient \citep[e.g.][]{2001ApJ...563..694V}. This means that substantial changes in the inclination-corrected linewidth of a galaxy are only expected if (i) the gas disc is severely truncated, such that most of the gas is at radii corresponding to the rising part of the rotation curve, or (ii) the total mass distribution changes such that the maximum circular velocity is significantly affected. To preserve the common shape of the \HIWF\ in the spring and fall skies, neither of these effects should be much stronger for galaxies in one sky region compared to the other.

We first consider what happens when a disc of \HI\ gas accretes additional gas, or is stripped or otherwise depleted. It is well established \citep[e.g.][]{1997AnA...324..877B,2001AnA...370..765V,2016MNRAS.460.2143W} that the \HI\ mass and diameter of galaxies are very tightly correlated. The authors of the latter reference point out that this implies that \HI\ discs grow/shrink in a well-regulated way as gas is accreted or consumed. This is further developed with the help of a series of analytic models by \citet{2019MNRAS.490...96S}, who find that galaxies likely evolve almost exactly along the correlation as their gas is consumed or stripped. \citet{2021MNRAS.502.5711N} further note that the $M_\mathrm{HI}$--$D_\mathrm{HI}$ correlation has no obvious connection to environment (though we feel this point could still be studied in more detail). The persistence of this very tight correlation in spite of even the messier aspects of galaxy evolution leads to a very useful point in the present discussion: if an \HI\ disc is truncated in radial extent, sufficiently that the portion of the rotation curve which it traces does not reach the outer, flat part, it must also necessarily be depleted in \HI\ mass, relative to its halo mass. Therefore, the less a galaxy's inclination-corrected 21-cm spectrum traces the circular velocity of its halo (due to limited extent of the \HI\ disc), the less likely it is to be detected in a flux-limited survey.

We next consider the conditions necessary to cause a substantial change to the maximum circular velocity of a galaxy. The most straightforward way for this to occur is via the stripping of large amounts of material. As the components which usually make the predominant contribution to the maximum circular velocity -- that is, dark matter, and for more massive galaxies, stars -- are collisionless, tidal stripping is the most important physical process in this context. By their nature, tides act to remove material approximately outside-in. For a typical galaxy, the maximum circular velocity actually does not change very much until a majority fraction of the total mass of the galaxy has been stripped \citep{2008ApJ...673..226P}; for example, a galaxy having lost $50$~per~cent of its total mass has its maximum circular velocity drop by only $\lesssim 15$~per~cent. Furthermore, usually no substantial amount of stellar mass is stripped until $\gtrsim 90$~per~cent of the dark matter has already been lost. The maximum circular velocity is therefore nearly a constant until rather late in the evolution of a satellite galaxy. In the meantime, there has likely been ample opportunity for the neutral gas disc to be stripped by ram pressure, consumed (and not replenished) by star formation, expelled by supernova or AGN winds, or a combination of the above. Thus, those galaxies where the connection between the maximum circular velocity and the (theoretically idealised, e.g. `pre-infall') halo mass are again the least likely to be detected in a 21-cm survey.

\subsubsection{Utility as a constraint on cosmology}

Putting the above considerations together, we propose the following qualitative interpretation of the shapes of the \HIMF\ and \HIWF\ in the \AA\ spring and fall skies. We suggest that the spatial variation in the \HIMF\ reflects the relatively fragile nature of \HI\ in galaxies: the atomic gas content responds readily to the local environment \citep[e.g.][and references therein]{2020MNRAS.494.2090J}. Thus, the distribution of \HI-to-total masses encoded in the \HIMF\ reflects local features, such as the presence of gas-rich filaments in the broad region around the Virgo cluster driving up the low-mass slope of the \HIMF\ in the \AA\ spring sky, as suggested by \citetalias{2018MNRAS.477....2J}. We further suggest that this same fragility somewhat counter-intuitively makes the \HIWF\ less sensitive to local variations in the strength of `environmental processes', because in most cases the \HI\ disc may lose so much mass that it drops out of the survey, or be completely removed/destroyed, before $w_{50}$ begins to be severely affected. To put this concisely, we suggest that environmental processes remove galaxies from the sample contributing to the \HIWF\ more effectively than they re-shape the velocity width distritbution of a galaxy population.

This leaves one remaining puzzle: for the shape of the \HIWF\ to remain unaffected, the population of galaxies which are not \HI-rich enough to be detected in a survey at any distance should not be differently biased in $w_{50}$ in different survey regions. Such gas-poor galaxies tend to be (i) massive, in terms of stellar or total mass, (ii) satellites or (iii) both. Given this, it seems plausible that, within any reasonably large volume, the fraction of very \HI-poor galaxies could be close enough to a constant at any given $w_{50}$ to give rise to the observed spatial invariance of the \HIWF. However, further theoretical work to reinforce or refute this hypothesis would be valuable.

To summarise: we argue that the \HIWF\ is a naturally reasonably robust\footnote{Not to be misunderstood as `direct' or `ideal'.} tracer of the HMF of gas-rich galaxies because flux-limited surveys are intrinsically biased against detecting galaxies where $w_{50}$ is a poor tracer of the halo mass. The spatial invariance of the \HIWF\ therefore reflects that of the HMF, even when comparing regions of dissimilar average density, and even though the relative gas-richness of the same population (i.e. the \HIMF) differs across the same regions.

Provided the interpretation outlined above holds, the position of the \HIWF\ as a cosmologically interesting quantity is reinforced: future measurements \citep[e.g.][]{2020ApnSS.365..118K} will deliver what may turn out to be the most stringent constraints on the dark matter HMF available, albeit for a biased selection of galaxies. In our opinion, the most fruitful way to exploit this will be for models to first reproduce the observed \HI-bearing galaxy population in sufficient detail within a given cosmological context, then to forward-model (i.e. `predict') the \HIWF\ for comparison with actual measurements. This is an approach we plan to pursue in future work.

\section{Conclusions}
\label{SecConc}

We have presented the first measurement of the \HIWF\ using the completed \AA\ survey ($\alpha.100$; Sec.~\ref{SecResults}). While the shape of the \HIMF\ is significantly different when the spring and fall portions of the survey are compared, the shape of the \HIWF\ in the same two regions is indistinguishable within the uncertainties, although these remain relatively large (Sec.~\ref{SecResults} \& \ref{SubsecSys}). We tentatively interpret this as a signature of the fidelity of the \HI\ line width as a tracer of the dynamical masses of galaxies (Sec.~\ref{SubsecInterpret}). We have also identified a previously oft-overlooked systematic bias particularly affecting the low-velocity slope of the \HIWF\ (Sec.\ref{SubsubsecUnseen}).

The scenario outlined in Sec.~\ref{SubsecInterpret} implies a few predictions; it will be interesting to see if these are borne out in future, larger-volume 21-cm surveys such as WALLABY \citep{2020ApnSS.365..118K} or an eventual SKA survey \citep[e.g.][]{2015aska.confE.128B}. First and foremost, the spatial invariance of the shape of the \HIWF\ can be confirmed or refuted as the statistical uncertainties shrink. Furthermore, we predict that the \HIWF\ should only maintain its shape when reasonably large (but not necessarily `cosmologically representative'), contiguous regions are compared. For example, the \HIMF\ measured in a collection of spatially disconnected voids would be expected to drive a strong bias against massive DM haloes, and therefore drive a reduction in $w_\star$ (and likely also some change in $\alpha$ and $\beta$). Finally, more sensitive surveys should uncover a previously unseen population of low-mass, high-velocity width galaxies, alleviating the bias discussed in Sec.~\ref{SubsubsecUnseen}. Another interesting possibility expected to be enabled by future surveys is the measurement of an \HI\ velocity function leveraging large numbers of spatially-resolved \HI\ sources to break the conventional reliance on ancillary optical data to estimate inclinations, or even replace line width measures with rotation velocities derived from kinematic models for thousands of sources. We look forward to harnessing the \AA\ and future \HIWF\ measurements as constraints on galaxy formation and cosmological models.

\section*{Acknowledgements}\label{sec-acknowledgements}

We are endebted to E.~Papastergis for providing us his implementation of the $1/V_\mathrm{eff}$ estimator and accompanying notes. We thank M.~Jones for providing tabulated results from \citetalias{2018MNRAS.477....2J} in electronic form. We thank both of the above, K.~Hess and A.~Ben\'{i}tez-Llambay for careful readings of draft versions of this article.

KAO acknowledges support by the European Research Council (ERC) through Advanced Investigator grant to C.S. Frenk, DMIDAS (GA 786910). This research has made use of NASA's Astrophysics Data System.

\section*{Data availability}

The $\alpha.100$ data release of the \AA\ survey, including the optical counterparts from \citet{2020AJ....160..271D}, is available at \url{egg.astro.cornell.edu/alfalfa/data/}.

\bibliography{paper}

\appendix

\section{Completeness of the \AA\ catalogues}
\label{AppComp}
The completeness limits for the \AA\ survey presented in \citetalias{2011AJ....142..170H} (eqs. 4 \& 5) are derived empirically based on the $\alpha.40$ catalogue. Although in principle the fully completed survey should have the same completeness as the 40~per~cent completed survey, since the former simply extends the sky coverage of the latter, the $\alpha.100$ catalogue offers the opportunity to both check this, and use the additional data to make a more precise determination of the limit. We have therefore repeated the calculation described in \citetalias{2011AJ....142..170H} (sec.~6) using the $\alpha.100$ `Code 1' sources. We also repeated the calculation using the $\alpha.40$ `Code 1' sources to verify that we could reproduce their result. We find a that the completeness limit implied by the $\alpha.40$ catalogue is in essentially the same location as reported by \citetalias{2011AJ....142..170H}, though we find a somewhat steeper cutoff as a function of flux (i.e. the spacings between the $90$, $50$ and $25$~per~cent completeness curves are somewhat narrower) -- we tentatively attribute this to small differences in the optimisation routines used to fit the number counts and the completeness curves themselves. For completeness, we include our determination of the limits based on the $\alpha.40$ catalogue here, following the notation of \citetalias{2011AJ....142..170H} for the left-hand sides of the equations:
\begin{align}
  \log_{10}S_\mathrm{21,90\%,Code1}&=
  \begin{cases}
    0.5W - 1.155, &W < 2.5 \\
    W - 2.405, &W \geq 2.5
  \end{cases}\\
  \log_{10}S_\mathrm{21,50\%,Code1}&=
  \begin{cases}
    0.5W - 1.194, &W < 2.5 \\
    W - 2.444, &W \geq 2.5
  \end{cases}\\
  \log_{10}S_\mathrm{21,25\%,Code1}&=
  \begin{cases}
    0.5W - 1.214, &W < 2.5 \\
    W - 2.464, &W \geq 2.5,
  \end{cases}
\end{align}
where $W=\log_{10}(w_{50}/\mathrm{km}\,\mathrm{s}^{-1})$. Using the $\alpha.100$ catalogue, we derive the following completeness limits:
\begin{align}
  \log_{10}S_\mathrm{21,90\%,Code1}&=
  \begin{cases}
    0.5W - 1.115, &W < 2.5 \\
    W - 2.365, &W \geq 2.5
  \end{cases}\\
  \log_{10}S_\mathrm{21,50\%,Code1}&=
  \begin{cases}
    0.5W - 1.170, &W < 2.5 \\
    W - 2.420, &W \geq 2.5
  \end{cases}\label{EqComp}\\
  \log_{10}S_\mathrm{21,25\%,Code1}&=
  \begin{cases}
    0.5W - 1.198, &W < 2.5 \\
    W - 2.248, &W \geq 2.5.
  \end{cases}
\end{align}
Note that these limits imply a completeness slightly ($\sim 0.02$~dex in $S_{21}$) worse than what is implied by the $\alpha.40$ catalogue. We use the $50$~per~cent completeness limit in Eq.~\ref{EqComp} throughout this work, except where explicitly noted otherwise.

\section{Reproduction of previously published results}
\label{AppPreviousResults}
We can reproduce the \HIWF\ measurement of \citetalias{2015AnA...574A.113P} exactly by using the public $\alpha.40$ galaxy catalogue and the selection cuts listed in that work, but this requires three adjustments. First, \citetalias{2015AnA...574A.113P} used a lower precision table of distances, so one object (AGC~7068) must be removed from the sample as its distance ($7.03\,\mathrm{Mpc}$) fails the \citetalias{2015AnA...574A.113P} distance cut ($D>7.0\,\mathrm{Mpc}$) at lower precision. Second, the original \citetalias{2015AnA...574A.113P} implementation of binning in $M_{\rm HI}$ and $w_{50}$ was subject to floating point roundoff errors, causing some objects on a bin boundary to be counted in the wrong bins, leading to small changes in the \HIWF. We have corrected this error, but verified that adopting the original (erroneous) binning scheme results in an exact quantitative reproduction of their result. Third, we must revert to the completeness limit given in \citetalias{2011AJ....142..170H} (eqs.~4\& 5). We also recover a near reproduction of the \citetalias{2015AnA...574A.113P} \HIWF\ by beginning with the $\alpha.100$ public catalogue, trimmed back to the $\alpha.40$ spring footprint (the region used by \citetalias{2015AnA...574A.113P}), and again applying the same selection cuts and binning as \citetalias{2015AnA...574A.113P}, and the \citetalias{2011AJ....142..170H} completeness limit -- this measurement is shown with light gray points in the upper left panel of Fig.~\ref{fig:hiwf}. The difference between the two measurements is due to the two reasons noted above, as well as two sources (AGC~749309, AGC~257959) being re-classified as OH megamasers in $\alpha.100$ \citep{2018ApJ...861...49H}.

The bulk of the difference between our measurement of the \HIMF\ and that reported in \citetalias{2018MNRAS.477....2J} is due to our use of an updated completeness limit for $\alpha.100$ (Eq.~\ref{EqComp}). Other small quantitative differences are due to: (i) the removal of a few sources re-classified as OH megamasers between the publication of \citetalias{2018MNRAS.477....2J} and the $\alpha.100$ catalogue release \citep{2018ApJ...861...49H}; (ii) our adjusted distances for a few objects (see Sec.~\ref{SecData}); (iii) the fact that we do not impose a minimum $M_\mathrm{HI}$ (rather than $10^{6}\,\mathrm{M}_\odot$); (iv) our use of the $1/V_\mathrm{eff}$ estimator, rather than the nearly-equivalent 2DSWML estimator.

\section{Tabulated \HIWF\ and \HIMF}
\label{AppTables}

In Tables~\ref{tab:wpoints} and \ref{tab:mpoints} we tabulate our \HIWF\ and \HIMF\ measurements, i.e. including amongst others all values plotted in the right panels of Fig.~\ref{fig:hiwf}.

\begin{table*}
  \caption{Amplitudes and uncertainties of the \HIWF\ for the full $\alpha.100$ survey (columns 2-3), and the spring and fall regions separately (columns 4-5 and 6-7, respectively). For each region, we give both the measurement including only counting uncertainties (left column of each pair) or all statistical uncertainties (right column of each pair), see Sec.~\ref{SecResults} for details of uncertainty estimates. Note that although the logarithms of all amplitudes and uncertainties are given, the counting uncertainties are symmetric on a linear scale and should be interpreted as the $1\sigma$ width of a Gaussian distribution (not a log-normal distribution). The intervals including all statistical uncertainties run from the $16^\mathrm{th}$ to $84^\mathrm{th}$ percentiles of the average of the $1000$ probability distributions each corresponding to a Monte Carlo-resampled galaxy catalogue.\label{tab:wpoints}}
  \begin{tabular}{ccccccc}
    \hline
    & \multicolumn{6}{c}{$\log_{10}\phi(w_{50})/h_{70}^3\,\mathrm{Mpc}^{-3}\,\mathrm{dex}^{-1}$} \\
    & \multicolumn{2}{c}{$\alpha.100$} & \multicolumn{2}{c}{$\alpha.100$ Spring} & \multicolumn{2}{c}{$\alpha.100$ Fall}\\
    $\log_{10}w_{50}/{\rm km}\,{\rm s}^{-1}$ & count. & all stat. & count. & all stat. & count. & all stat. \\
    \hline
    \input{tables/appendixtable_WF.tex}
    \hline
  \end{tabular}
\end{table*}

\begin{table*}
  \caption{As Table~\ref{tab:wpoints}, but for the \HIMF.\label{tab:mpoints}}
  \begin{tabular}{ccccccc}
    \hline
    & \multicolumn{6}{c}{$\log_{10}\phi(M_\mathrm{HI})/h_{70}^3\,\mathrm{Mpc}^{-3}\,\mathrm{dex}^{-1}$} \\
    & \multicolumn{2}{c}{$\alpha.100$} & \multicolumn{2}{c}{$\alpha.100$ Spring} & \multicolumn{2}{c}{$\alpha.100$ Fall}\\
    $\log_{10}M_\mathrm{HI}/{\rm M}_\odot$ & count. & all stat. & count. & all stat. & count. & all stat. \\
    \hline
    \input{tables/appendixtable_MF.tex}
    \hline
  \end{tabular}
\end{table*}

\section{Marginalised posterior probability distributions of Schechter function parameters}
\label{AppMFCorner}
\begin{figure*}
  \includegraphics[width=\textwidth]{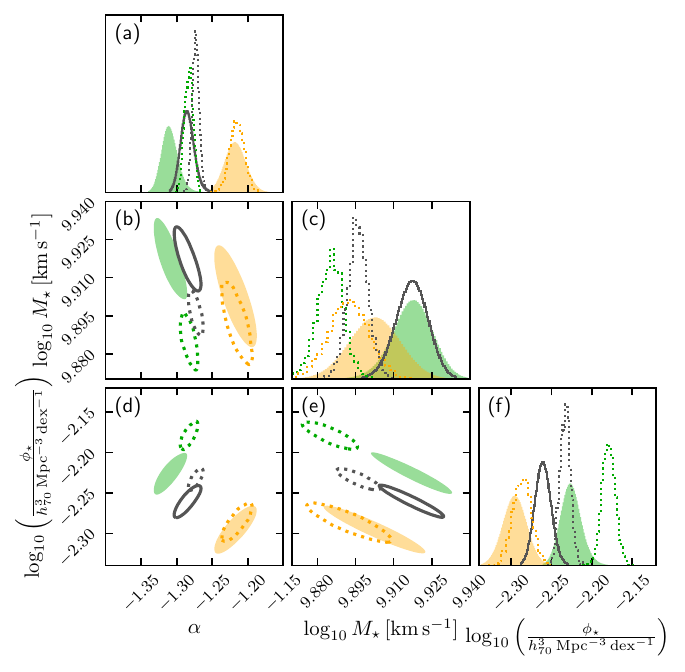}
  \caption{One- and two-dimensional marginalised posterior probability distributions for the parameters of the Schechter function (Eq.~\ref{eq:schechter}). All contours are drawn at the $95$~per~cent confidence level. Dark grey dotted contours and histograms correspond to the fit to the \HIMF\ of the full $\alpha.100$ sample when only counting (Poisson) uncertainties are accounted for. Dark grey solid contours and histograms are similar, but include all statistical uncertainties. Dark green and orange contours and dotted histograms show the parameter constraints for the spring and fall sky areas, respectively, accounting only for counting uncertainties, while the light green and orange contours and filled histograms show the constraints when all statistical uncertainties are accounted for. The difference in the shape parameters $(\alpha, M_\star)$ between the spring and fall regions noted by \citetalias{2018MNRAS.477....2J} is clearly visible in panel (b).\label{fig:corner_mf}}
\end{figure*}

We show the marginalised posterior probability distribution for the Schechter function parameters describing the \HIMF\ in Fig.~\ref{fig:corner_mf}, analogously to those for the \HIWF\ shown in Fig.~\ref{fig:corner}. The same approach as for the \HIWF\ is used to estimate the confidence regions including all statistical uncertainties -- see Sec.~\ref{SecResults} for details.

\section{Mock surveys}
\label{AppMocks}

We wish to construct a mock catalogue of `observed' sources drawn from a known underlying `intrinsic' source population. The distribution of \HI\ masses of the intrinsic population should be described by a Schechter function, and the velocity width distribution by a modified Schechter function. It is trivial to draw a set of discrete values from the two desired distributions to achieve this, but once this is done satisfying the additional desired constraint that the observed source population resemble that in the \AA\ survey is not trivial, and we could not arrive at a satisfactory formulation of the problem that is mathematically well-posed and not under-constrained. We therefore opted for a different approach. Since we are most interested in the \HIWF, we sampled a set of $w_{50}$ values from the desired distribution. We then assumed a Gumbel distribution (also known as a Type~I Extreme Value distribution) form for the distribution of \HI\ masses at fixed $w_{50}$ and that the two parameters describing the distribution, a width $\beta$ and centroid $\mu$, vary continuously with $w_{50}$. We iteratively adjusted the scalings of $\mu$ and $\beta$ brute-force `by hand' until (i) the amplitude of the resulting \HIMF\ matched that desired within 10~per~cent in each bin (the bins used are the same as those used in the $1/V_\mathrm{eff}$ estimator), and (ii) the observed source distribution in the $M_\mathrm{HI}$--$w_{50}$ plane was an acceptable match to the \AA\ counts. Criterion (ii) was assessed subjectively, `by eye' -- in combination with constraint (i) requiring a quantitative match proved intractable.

Once an \HI\ mass was assigned to each line width, the sources were distributed random-uniformly in a survey volume and their 21-cm fluxes computed from their distances and \HI\ masses. All sources in this intrinsic source list below the \AA\ 50~per~cent completeness limit were removed to yield the `observed' source catalogue.

The Gumbel distribution parameters used to construct the mocks are as follows, where $M=\log_{10}(M_\mathrm{HI}/\mathrm{M}_\odot)$ and $W=\log_{10}(w_{50}/{\rm km}\,{\rm s}^{-1})$. For mock~(a):
\begin{align}
  \mu(W) = \begin{cases}
    5.47W - 2.22 &\text{if $1.20\leq W<1.76$} \\
    3.11W + 1.92 &\text{if $1.76\leq W<2.20$} \\
    2.38W + 3.55 &\text{if $2.20\leq W<3.00$}
  \end{cases} \\
  \beta(W) = \begin{cases}
    -1.00W + 2.90 &\text{if $1.20\leq W<1.80$} \\
    -0.81W + 2.56 &\text{if $1.80\leq W<2.60$} \\
    -0.68W + 2.21 &\text{if $2.60\leq W<3.00$} \\
  \end{cases},
\end{align}
for mock~(b):
\begin{align}
  \mu(W) = \begin{cases}
    6.51W - 3.86 &\text{if $1.20\leq W<1.76$} \\
    3.20W + 1.96 &\text{if $1.76\leq W<2.20$} \\
    2.21W + 4.14 &\text{if $2.20\leq W<3.00$}
  \end{cases} \\
  \beta(W) = \begin{cases}
    -0.34W + 2.21 &\text{if $1.20\leq W<1.40$} \\
    -1.94W + 4.45 &\text{if $1.40\leq W<1.76$} \\
    -1.00W + 2.79 &\text{if $1.76\leq W<2.20$} \\
    -0.53W + 1.75 &\text{if $2.20\leq W<3.00$} \\
  \end{cases},
\end{align}
for mock~(c):
\begin{align}
  \mu(W) = \begin{cases}
    6.05W - 2.44 &\text{if $1.20\leq W<1.76$} \\
    2.95W + 3.00 &\text{if $1.76\leq W<2.20$} \\
    1.64W + 5.90 &\text{if $2.20\leq W<3.00$}
  \end{cases} \\
  \beta(W) = \begin{cases}
    0.4W + 0.94 &\text{if $1.20\leq W<1.40$} \\
    -2.08W + 4.42 &\text{if $1.40\leq W<1.76$} \\
    -0.80W + 2.15 &\text{if $1.76\leq W<2.20$} \\
    -0.31W + 1.09 &\text{if $2.20\leq W<3.00$} \\
  \end{cases},
\end{align}
and the cumulative distribution function of the Gumbel distribution is:
\begin{align}
  F(M|W) &= \frac{e^{-e^{-z(M,W)}}-e^{-e^{-z_\mathrm{min}}}}{e^{-e^{-z_\mathrm{max}}}-e^{-e^{-z_\mathrm{min}}}}\label{eq:CDF}
\end{align}
with
\begin{align}
  z(M,W) &= \frac{\mu(W)-M}{\beta(W)}, \\
  z_\mathrm{min} &= z(M_\mathrm{min}, W), \\
  z_\mathrm{max} &= z(M_\mathrm{max}, W)
\end{align}
where $M_\mathrm{min}$ and $M_\mathrm{max}$ correspond to the ($\log_{10}$ of the) minimum and maximum \HI\ masses to be sampled, in our case $6.0$ and $11.0$, respectively. To determine a mass for a mock source of velocity width $W$, one therefore simply draws a random value from the distribution given by Eq.~\ref{eq:CDF}.

\label{lastpage}
\end{document}

%% file: tables/appendixtable_WF.tex
$1.2$&$-0.23_{-0.35}^{+0.19}$&$-0.51_{-0.53}^{+0.33}$&$-0.30_{-0.33}^{+0.18}$&$-0.55_{-0.58}^{+0.44}$&$-0.44_{-0.98}^{+0.28}$&$-0.86_{-0.95}^{+0.62}$\\
$1.4$&$-0.19_{-0.23}^{+0.15}$&$-0.27_{-0.35}^{+0.27}$&$0.03_{-0.28}^{+0.17}$&$-0.13_{-0.42}^{+0.32}$&$-0.72_{-0.22}^{+0.15}$&$-0.80_{-0.45}^{+0.52}$\\
$1.5$&$-0.29_{-0.46}^{+0.22}$&$-0.46_{-0.27}^{+0.30}$&$0.04_{-0.85}^{+0.27}$&$-0.36_{-0.35}^{+0.47}$&$-0.72_{-0.20}^{+0.14}$&$-0.79_{-0.23}^{+0.24}$\\
$1.6$&$-0.73_{-0.07}^{+0.06}$&$-0.67_{-0.14}^{+0.24}$&$-0.63_{-0.09}^{+0.07}$&$-0.59_{-0.17}^{+0.30}$&$-0.91_{-0.11}^{+0.09}$&$-0.90_{-0.19}^{+0.23}$\\
$1.7$&$-0.91_{-0.05}^{+0.05}$&$-0.78_{-0.12}^{+0.20}$&$-0.79_{-0.06}^{+0.05}$&$-0.67_{-0.14}^{+0.26}$&$-1.14_{-0.11}^{+0.09}$&$-1.07_{-0.16}^{+0.18}$\\
$1.8$&$-0.99_{-0.04}^{+0.04}$&$-0.88_{-0.12}^{+0.22}$&$-0.91_{-0.05}^{+0.05}$&$-0.78_{-0.14}^{+0.26}$&$-1.11_{-0.08}^{+0.07}$&$-1.13_{-0.12}^{+0.12}$\\
$1.9$&$-1.00_{-0.05}^{+0.05}$&$-0.90_{-0.11}^{+0.19}$&$-0.89_{-0.07}^{+0.06}$&$-0.79_{-0.14}^{+0.24}$&$-1.22_{-0.07}^{+0.06}$&$-1.17_{-0.10}^{+0.10}$\\
$2.0$&$-0.93_{-0.05}^{+0.04}$&$-0.85_{-0.11}^{+0.20}$&$-0.80_{-0.06}^{+0.05}$&$-0.73_{-0.12}^{+0.23}$&$-1.26_{-0.05}^{+0.05}$&$-1.22_{-0.08}^{+0.10}$\\
$2.1$&$-0.99_{-0.05}^{+0.04}$&$-0.97_{-0.09}^{+0.17}$&$-0.88_{-0.06}^{+0.05}$&$-0.85_{-0.11}^{+0.23}$&$-1.25_{-0.07}^{+0.06}$&$-1.26_{-0.08}^{+0.10}$\\
$2.2$&$-1.15_{-0.04}^{+0.04}$&$-1.09_{-0.08}^{+0.17}$&$-1.03_{-0.05}^{+0.04}$&$-0.98_{-0.09}^{+0.20}$&$-1.45_{-0.04}^{+0.03}$&$-1.39_{-0.06}^{+0.09}$\\
$2.3$&$-1.29_{-0.03}^{+0.03}$&$-1.27_{-0.04}^{+0.07}$&$-1.20_{-0.04}^{+0.03}$&$-1.18_{-0.05}^{+0.08}$&$-1.47_{-0.04}^{+0.04}$&$-1.45_{-0.05}^{+0.06}$\\
$2.4$&$-1.55_{-0.03}^{+0.03}$&$-1.51_{-0.04}^{+0.05}$&$-1.47_{-0.03}^{+0.03}$&$-1.44_{-0.05}^{+0.06}$&$-1.71_{-0.03}^{+0.03}$&$-1.67_{-0.04}^{+0.05}$\\
$2.5$&$-1.64_{-0.04}^{+0.04}$&$-1.64_{-0.05}^{+0.06}$&$-1.55_{-0.05}^{+0.05}$&$-1.57_{-0.07}^{+0.07}$&$-1.89_{-0.03}^{+0.03}$&$-1.83_{-0.05}^{+0.05}$\\
$2.6$&$-1.96_{-0.04}^{+0.04}$&$-1.91_{-0.06}^{+0.07}$&$-1.88_{-0.05}^{+0.05}$&$-1.84_{-0.07}^{+0.08}$&$-2.10_{-0.07}^{+0.06}$&$-2.07_{-0.07}^{+0.07}$\\
$2.7$&$-2.12_{-0.10}^{+0.08}$&$-2.09_{-0.12}^{+0.18}$&$-2.00_{-0.11}^{+0.09}$&$-1.98_{-0.15}^{+0.23}$&$-2.50_{-0.07}^{+0.06}$&$-2.41_{-0.08}^{+0.10}$\\
$2.8$&$-2.96_{-0.07}^{+0.06}$&$-2.84_{-0.09}^{+0.11}$&$-2.88_{-0.10}^{+0.08}$&$-2.75_{-0.11}^{+0.12}$&$-3.08_{-0.07}^{+0.06}$&$-3.01_{-0.10}^{+0.12}$\\
$2.9$&$-4.20_{-0.18}^{+0.12}$&$-3.90_{-0.26}^{+0.29}$&$-4.06_{-0.25}^{+0.16}$&$-3.85_{-0.29}^{+0.31}$&$-4.44_{-0.25}^{+0.16}$&$-4.16_{-0.36}^{+0.53}$\\

%% file: tables/appendixtable_MF.tex
$6.1$&$-1.00_{-0.54}^{+0.23}$&$-1.14_{-0.67}^{+0.35}$&$-0.74_{-0.54}^{+0.23}$&$-0.92_{-0.72}^{+0.36}$&--&$-1.09_{-1.48}^{+0.33}$\\
$6.3$&$-1.56_{-15.90}^{+0.30}$&$-1.28_{-0.62}^{+0.36}$&--&$-1.12_{-0.69}^{+0.38}$&--&$-1.28_{-1.33}^{+0.36}$\\
$6.5$&$-1.21_{-0.30}^{+0.18}$&$-1.20_{-0.40}^{+0.26}$&$-1.07_{-0.38}^{+0.20}$&$-1.13_{-0.54}^{+0.32}$&--&$-1.40_{-0.70}^{+0.36}$\\
$6.7$&$-1.22_{-0.23}^{+0.15}$&$-1.19_{-0.26}^{+0.20}$&$-1.31_{-0.32}^{+0.18}$&$-1.10_{-0.34}^{+0.24}$&$-1.20_{-0.40}^{+0.20}$&$-1.39_{-0.52}^{+0.29}$\\
$6.9$&$-1.35_{-0.15}^{+0.11}$&$-1.14_{-0.16}^{+0.13}$&$-1.50_{-0.20}^{+0.14}$&$-1.02_{-0.20}^{+0.16}$&$-1.25_{-0.23}^{+0.15}$&$-1.41_{-0.32}^{+0.22}$\\
$7.1$&$-1.11_{-0.07}^{+0.06}$&$-1.11_{-0.11}^{+0.10}$&$-0.98_{-0.08}^{+0.07}$&$-0.98_{-0.13}^{+0.11}$&$-1.43_{-0.19}^{+0.13}$&$-1.47_{-0.26}^{+0.19}$\\
$7.3$&$-1.10_{-0.06}^{+0.05}$&$-1.11_{-0.08}^{+0.07}$&$-0.94_{-0.07}^{+0.06}$&$-0.97_{-0.09}^{+0.08}$&$-1.53_{-0.16}^{+0.12}$&$-1.47_{-0.21}^{+0.16}$\\
$7.5$&$-1.14_{-0.05}^{+0.04}$&$-1.19_{-0.06}^{+0.06}$&$-1.01_{-0.05}^{+0.05}$&$-1.09_{-0.07}^{+0.07}$&$-1.44_{-0.12}^{+0.10}$&$-1.44_{-0.16}^{+0.13}$\\
$7.7$&$-1.25_{-0.04}^{+0.04}$&$-1.28_{-0.05}^{+0.05}$&$-1.18_{-0.04}^{+0.04}$&$-1.21_{-0.06}^{+0.06}$&$-1.38_{-0.09}^{+0.07}$&$-1.40_{-0.12}^{+0.10}$\\
$7.9$&$-1.33_{-0.03}^{+0.03}$&$-1.35_{-0.04}^{+0.04}$&$-1.26_{-0.04}^{+0.04}$&$-1.28_{-0.05}^{+0.05}$&$-1.45_{-0.08}^{+0.07}$&$-1.49_{-0.10}^{+0.09}$\\
$8.1$&$-1.39_{-0.03}^{+0.03}$&$-1.41_{-0.04}^{+0.04}$&$-1.31_{-0.03}^{+0.03}$&$-1.33_{-0.04}^{+0.04}$&$-1.56_{-0.07}^{+0.06}$&$-1.62_{-0.09}^{+0.08}$\\
$8.3$&$-1.43_{-0.03}^{+0.02}$&$-1.44_{-0.03}^{+0.03}$&$-1.32_{-0.03}^{+0.03}$&$-1.35_{-0.04}^{+0.03}$&$-1.69_{-0.06}^{+0.05}$&$-1.65_{-0.07}^{+0.06}$\\
$8.5$&$-1.49_{-0.02}^{+0.02}$&$-1.51_{-0.03}^{+0.03}$&$-1.39_{-0.03}^{+0.03}$&$-1.42_{-0.03}^{+0.03}$&$-1.67_{-0.04}^{+0.04}$&$-1.69_{-0.06}^{+0.05}$\\
$8.7$&$-1.55_{-0.02}^{+0.02}$&$-1.57_{-0.02}^{+0.02}$&$-1.47_{-0.02}^{+0.02}$&$-1.50_{-0.03}^{+0.03}$&$-1.68_{-0.03}^{+0.03}$&$-1.68_{-0.04}^{+0.04}$\\
$8.9$&$-1.64_{-0.02}^{+0.02}$&$-1.63_{-0.02}^{+0.02}$&$-1.56_{-0.02}^{+0.02}$&$-1.57_{-0.03}^{+0.02}$&$-1.75_{-0.03}^{+0.02}$&$-1.72_{-0.03}^{+0.03}$\\
$9.1$&$-1.68_{-0.01}^{+0.01}$&$-1.68_{-0.01}^{+0.01}$&$-1.61_{-0.02}^{+0.02}$&$-1.63_{-0.02}^{+0.02}$&$-1.76_{-0.02}^{+0.02}$&$-1.76_{-0.02}^{+0.02}$\\
$9.3$&$-1.79_{-0.01}^{+0.01}$&$-1.79_{-0.01}^{+0.01}$&$-1.73_{-0.01}^{+0.01}$&$-1.74_{-0.02}^{+0.02}$&$-1.88_{-0.01}^{+0.01}$&$-1.88_{-0.02}^{+0.02}$\\
$9.5$&$-1.95_{-0.01}^{+0.01}$&$-1.96_{-0.01}^{+0.01}$&$-1.90_{-0.01}^{+0.01}$&$-1.91_{-0.01}^{+0.01}$&$-2.03_{-0.01}^{+0.01}$&$-2.03_{-0.02}^{+0.02}$\\
$9.7$&$-2.11_{-0.01}^{+0.01}$&$-2.12_{-0.01}^{+0.01}$&$-2.08_{-0.01}^{+0.01}$&$-2.08_{-0.01}^{+0.01}$&$-2.17_{-0.01}^{+0.01}$&$-2.18_{-0.02}^{+0.01}$\\
$9.9$&$-2.32_{-0.01}^{+0.01}$&$-2.32_{-0.01}^{+0.01}$&$-2.28_{-0.01}^{+0.01}$&$-2.28_{-0.01}^{+0.01}$&$-2.39_{-0.01}^{+0.01}$&$-2.39_{-0.01}^{+0.01}$\\
$10.1$&$-2.62_{-0.01}^{+0.01}$&$-2.61_{-0.01}^{+0.01}$&$-2.58_{-0.01}^{+0.01}$&$-2.57_{-0.01}^{+0.01}$&$-2.67_{-0.01}^{+0.01}$&$-2.66_{-0.02}^{+0.02}$\\
$10.3$&$-3.07_{-0.01}^{+0.01}$&$-3.05_{-0.01}^{+0.01}$&$-3.06_{-0.02}^{+0.02}$&$-3.03_{-0.02}^{+0.02}$&$-3.09_{-0.02}^{+0.02}$&$-3.07_{-0.02}^{+0.02}$\\
$10.5$&$-3.77_{-0.03}^{+0.03}$&$-3.73_{-0.03}^{+0.03}$&$-3.74_{-0.04}^{+0.03}$&$-3.70_{-0.04}^{+0.04}$&$-3.83_{-0.05}^{+0.05}$&$-3.78_{-0.06}^{+0.05}$\\
$10.7$&$-4.64_{-0.09}^{+0.07}$&$-4.61_{-0.09}^{+0.08}$&$-4.63_{-0.11}^{+0.09}$&$-4.59_{-0.11}^{+0.09}$&$-4.65_{-0.16}^{+0.11}$&$-4.66_{-0.17}^{+0.13}$\\
$10.9$&$-5.81_{-0.53}^{+0.23}$&$-5.74_{-0.47}^{+0.23}$&$-5.61_{-0.53}^{+0.23}$&$-5.59_{-0.52}^{+0.23}$&--&$-5.69_{-1.99}^{+0.30}$\\